\documentclass[letterpaper, 11pt]{article}
\pdfoutput = 1

\usepackage{setspace}
\usepackage[bookmarks = false, colorlinks = true, linkcolor = blue, citecolor = purple]{hyperref}
\usepackage{shortcuts}
\usepackage{titlesec}
\usepackage{cite}
\usepackage{tocloft}
\usepackage[vcentermath]{youngtab}
\usepackage{tensor}

\usepackage{tikz}
\usetikzlibrary{arrows, decorations.pathmorphing, decorations.markings, backgrounds, positioning, fit, petri}

\usepackage[outline]{contour}
\contourlength{1.2pt}

\usepackage[margin = 2.5cm]{geometry}
\begin{document}

\thispagestyle{empty}

\begin{center}

\vspace*{5em}

{\LARGE \bf Spin Structures and Exact Dualities in Low Dimensions}

\vspace{1cm}

{\large \DJ or\dj e Radi\v cevi\'c}
\vspace{1em}

{\it Perimeter Institute for Theoretical Physics, Waterloo, Ontario, Canada N2L 2Y5}\\
\texttt{djordje@pitp.ca}\\

\vspace{0.08\textheight}
\begin{abstract}
This paper derives a large web of exact lattice dualities in one and two spatial dimensions. Some of the dualities are well-known, while others, such as two-dimensional boson-parafermion dualities, are new. The procedure is systematic, independent of specific Hamiltonians, and generalizes to higher dimensions. One important result is a demonstration that spin structures in arbitrary lattice fermion theories can always be simply defined as topological gauge fields whose gauge group is the fermion number parity. This definition agrees with other expected properties of spin structures, and it motivates the introduction of ``paraspin structures'' that serve the same role in parafermion theories.
\end{abstract}
\end{center}

\newpage

\tableofcontents

\section{Introduction}

Dualities are ubiquitous in quantum physics. Distinguished among them are \emph{exact dualities}, in which every state and operator --- at every energy scale --- has a known map between two theories. Exact dualities can be \emph{proven} for entire classes of theories, and they are among the most powerful tools available for studying high-energy, strongly quantum regimes. Perhaps the best example of such an application is the solution of the 2D Ising model at all couplings via exact fermionization. Other important exact dualities are 4D Abelian electric-magnetic duality, 3D particle-vortex duality (in its ``weak'' version \cite{Senthil:2018}), and 3D Chern-Simons level-rank dualities.

Unsurprisingly, such power comes with a price: in order to prove exact dualities, a fully regularized, nonperturbative formulation of a theory must be known. Lattice theories (e.g.~\cite{Wilson:1974sk}) and axiomatically defined topological field theories \cite{Atiyah:1988} are standard examples of these formulations. At this time, many theories of interest do not appear to naturally lend themselves to such regularization, the most flagrant examples being the Standard Model (and other chiral theories), nonlagrangian quantum field theories, and most theories of gravity in more than two dimensions. Nevertheless, studying exact dualities in the situations where they may be proven already provides a wealth of insights into quantum field theory.

This paper deals with exact dualities in Hamiltonian lattice systems in one and two spatial dimensions.%
\footnote{The focus on low dimensions is more a matter of convenience: this way it is much simpler to illustrate the main ideas. All dualities discussed in this paper will have a straightforward (if notationally horrendous) generalization to higher dimensions, and occasionally these generalizations will be spelled out in this work. \\ %
\indent The focus on Hamiltonian (operator) methods is chosen for three reasons. One is, again, convenience: instead of working in $D$ spacetime dimensions, one can work in $d = D - 1$ spatial dimensions where the geometry is simpler. Another reason is the generality: when working with operator algebras, no particular Hamiltonian (or action) needs to be specified when proving exact dualities. Thus the requirements for a duality to hold become more transparent,  as does the counting of degrees of freedom. The final reason is related to this last point: operator methods make it clear how to define entanglement entropy and thus how to study the entanglement structure of states even when the Hilbert space does not factorize, for instance due to the presence of gauge constraints. While entanglement will not be the subject of this paper, the analysis presented here synergizes with that in \cite{Lin:2018bud} to make it possible to study, for example, entanglement of spin structures or the detailed mapping of entanglement under exact dualities.} %
It has multiple goals. The first, rather down-to-earth goal is to generalize the recently established 3D bosonization duality \cite{Chen:2017fvr} to spatial surfaces of higher genus and to parafermions. The second goal is to compile a mini-compendium of exact lattice dualities with all the important details spelled out, extending and streamlining the overview \cite{Cobanera:2011wn}. The third goal is to systematize various notions of ``twisting'' that can be performed on dual pairs of theories in order to derive new dualities.%
\footnote{This is related to promoting a background gauge field to a dynamical one. A familiar example is the $S$ generator of the Kapustin-Strassler-Witten SL$(2,\Z)$ group action on 3D theories with U(1) symmetry \cite{Kapustin:1999ha, Witten:2003ya}, which can be used to elegantly create duality webs in three dimensions. In this paper, an analogous procedure is applied to theories with any Abelian symmetry in any dimension.} %
The fourth and perhaps most important goal is to clarify what it means for a generic (possibly nonrelativistic and nontopological) fermion theory to depend on a spin structure.

This final goal deserves some further elaboration. Spin structures routinely appear in relativistic fermionic theories (see e.g.~\cite{Seiberg:1986by, AlvarezGaume:1987vm}), but they exist in discrete theories, too (see e.g.~\cite{Bhardwaj:2016clt, Cimasoni:2008, Gaiotto:2015zta, Kapustin:2017jrc, Tarantino:2016qfy, Wang:2017moj, Ellison:2018}). How to define spin structures in a general way? One important lesson of this paper is that spin structures can always be viewed as $\Z_2$ topological gauge fields whose gauge group is generated by the fermion number parity operator $(-1)^F$.\footnote{In one spatial dimension, this story can be made more intricate due to various low-dimensional coincidences that will be discussed throughout the main text.} The distinction between \emph{spin} and \emph{non-spin} theories, i.e.~between theories that do and do not depend on a choice of spin structure, is equivalent to the distinction between spin structures being background or dynamical fields. In both cases, these gauge fields can be called topological because their field strength is fixed by requiring that they encode all ordering ambiguities of the fermionic Hilbert space: this field strength can be canonically chosen to equal the second Stiefel-Whitney class of the manifold in question. This point of view permeates many papers, in particular \cite{Bhardwaj:2016clt}, but appears not to have been articulated in simple terms. This paper aims to fill this gap.

The rest of this work is organized in a relatively straightforward way: exact dualities are examined systematically, for $d = 1$ and $d = 2$ spatial dimensions, for boson-boson (Kramers-Wannier \cite{Kramers:1941}), boson-fermion (Jordan-Wigner \cite{Jordan:1928}), and boson-parafermion (Fradkin-Kadanoff \cite{Fradkin:1980th}) maps, and for all natural twists and generalizations that present themselves along the way. The bulk of the text is a derivation of various dualities, most of which are summarized below for covenience. (Checks (${}^\vee$) indicate that the theory lives on a dual lattice, and ``/G'' denotes projection to the singlet sector of a global symmetry G, which may be ordinary or higher-form \cite{Gaiotto:2014kfa}. Other details of various dualities are explained in the corresponding sections of the main text.) The end of each section serves to draw more general lessons about dualities discussed up to that point.

\begin{center}
\begin{tabular}{ccc}
  \textbf{Name} & \textbf{Duality} & \textbf{Eq.}\\ \hline
  $d = 1$ Kramers-Wannier & spins$/\Z_2$ = spins$^\vee/\Z_2$  & \eqref{1d KW original} \\
  & spins = gauged spins$^\vee$ & \eqref{1d KW alternative} \\
  & $\Z_K$ clock model/$\Z_K$ = $\Z_K$ clock model$^\vee/\Z_K$ & \eqref{1d KW Zk}\\
  & $q$-gauged $\Z_K$ model/$\Z_q$ = $\Z_q$-orbifolded $\Z_{K/q}$ model$^\vee$ & \eqref{1d KW Zk orbifold} \\
  particle-kink & compact scalar/U(1) = compact scalar$^\vee$/U(1) & \eqref{1d KW Zk cont}\\
  & compact scalar = gauged compact scalar$^\vee$ & \eqref{1d KW Zk alt cont}\\
  $d = 1$ Jordan-Wigner & spins = fermions & \eqref{1d JW original} \\
  & spins$/\Z_2$ = fermions$/\Z_2$ & \eqref{1d JW restricted}\\
  & spins = $\Z_2$ QED$^\vee$ & \eqref{1d JW alternative}\\
  $d = 1$ Fradkin-Kadanoff & $\Z_K$ clock model = $\Z_K$ parafermions & \eqref{1d FK original}\\ \hline %
  $d = 2$ Kramers-Wannier & spins$/\Z_2$ = $\Z_2$ gauge theory$^\vee/(\Z_2$ one-form) & \eqref{2d KW original}\\
  & topologically gauged spins = $\Z_2$ gauge theory$^\vee$  & \eqref{2d KW alternative}\\
  & spins = gauged $\Z_2$ gauge theory$^\vee$ & \eqref{2d KW alternative 2}\\
  & $\Z_K$ clock model/$\Z_K$ = $\Z_K$ gauge theory$^\vee/(\Z_K$ one-form) & \eqref{2d KW Zk original}\\
  particle-vortex & compact scalar/U(1) = U(1) g.~theory$^\vee$/(U(1) one-form) & \eqref{2d KW cont} \\
  $d = 2$ Jordan-Wigner & fermions/$\Z_2$ = flux-attached $\Z_2$ g.~theory$^\vee/(\Z_2$ one-form) & \eqref{2d JW twisted}\\
  & topologically gauged fermions = flux-attached $\Z_2$ g.~theory$^\vee$ & \eqref{2d JW alternative}\\
  $d = 2$ Fradkin-Kadanoff & $\Z_K$ paraferms./$\Z_K$ = flux-att.~$\Z_K$ g.~theory$^\vee/(\Z_K$ one-form) & \eqref{2d FK} \\
  & $\frac qK$ paraferms.$/\Z_K$ = $q$-flux-att.~$\Z_K$ g.~theory$^\vee$/($\Z_K$ one-form)& \eqref{2d FK any} \\ \hline
  & comp.\ scalar/U(1) = $(d - 1)$-form g.~theory$^\vee$/(U(1) one-form) & \eqref{cont duality 1} \\
  (any $d$, boson-boson) & topologically gauged comp.\ scalar = $(d - 1)$-form g.~theory$^\vee$ & \eqref{cont duality 2} \\
  & comp.\ scalar = gauged $(d - 1)$-form g.~theory$^\vee$ & \eqref{cont duality 3}\\ \hline
\end{tabular}
\end{center}

\section{$\Z_2$ dualities in $d = 1$}

\begin{figure}[b!]
\begin{center}

\begin{tikzpicture}[scale = 2]
  \foreach \i in {1,...,9} {
    \coordinate (N\i) at (\i*360/9:1); %need \i * 360/(# points) : radius
    \filldraw[draw = black, fill = black] (N\i) circle (0.05);
  }
  \draw[red] (0, 0) circle (0.975);
  \foreach \i in {1,...,9} {
    \coordinate (N\i) at ({(\i*360+180)/9}:0.975); %need \i * 360/(# points) : radius
    \filldraw[draw = red, fill = white] (N\i) circle (0.05);
  }
  \draw (0, 0) circle (1);
\end{tikzpicture}
\end{center}
\caption{\small A periodic lattice with $N = 9$ sites (black) and $N$ edges between them. In this section, each site $v$ of such a lattice hosts a two-dimensional Hilbert space of an Ising spin, and the algebra of operators at that site is generated by the Pauli matrices $X_v$ and $Z_v$.  The dual (red) lattice has a site corresponding to each of the edges of the original lattice, and an edge corresponding to each vertex of the dual lattice. The edge-vertex duality is the $d = 1$ avatar of the more general Poincar\'e duality.}
\label{fig lattices}
\end{figure}
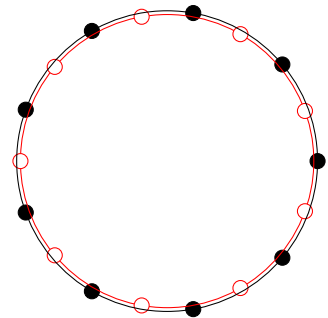

Even though scalar field theories are canonical starting points for quantum field theorists, it is more illuminating to start the overview of dualities with an even simpler setup and to generalize from there. Consider a system of bosonic $\Z_2$ degrees of freedom (``Ising spins'' or just ``spins'') on $N$ sites arranged in a circle (fig.~\ref{fig lattices}). Its algebra of operators is generated by pairs of Pauli matrices $X_v$ and $Z_v$ on each site $v$. In the $Z$-eigenbasis, the operators on each site can be recorded as
\bel{
  Z = \bmat100{-1}, \quad X = \bmat0110.
}
Unsurprisingly, the archetypical theory of Ising spins is the Ising model in a transverse field $h$, given by the Hamiltonian
\bel{\label{def Ising}
  H\_{Ising} = \sum_{v = 1}^N \left(X_v X_{v + 1} + h Z_v\right), \quad  X_{N + 1} \equiv X_1.
}

The transverse field Ising model --- and essentially any other theory where the operator algebra is generated by $\{X_v, Z_v\}$ --- exhibits two well-known classes of exact dualities: Kramers-Wannier (boson-boson, \cite{Kramers:1941}) and Jordan-Wigner (boson-fermion, \cite{Jordan:1928}). Subsections \ref{subsec 1d KW} and \ref{subsec 1d JW} are dedicated to reviewing these standard ideas and explaining the word ``essentially'' in the previous sentence. A systematic procedure called \emph{duality twisting} will then be described in subsection \ref{subsec twisting}, where it will be used to establish a web of $d = 1$ dualities in a way that can be generalized to higher dimensions.

\subsection{Kramers-Wannier} \label{subsec 1d KW}

Kramers-Wannier (KW) duality maps Ising spins to Ising spins on the dual lattice (also shown on fig.~\ref{fig lattices}). The dual lattice is again a circle of $N$ sites which can be conveniently labeled by half-integers $\frac12, \frac32, \ldots, N - \frac12$, with dual site $v + \frac12$ corresponding to the edge between sites $v$ and $v + 1$. The standard version of the KW mapping is
\algns{\label{1d KW original}
  Z_v &= X^\vee_{v - \frac12} X^\vee_{v + \frac 12}, \\
  X_v X_{v + 1} &= Z^\vee_{v + \frac12}.
}
The index $v$ here takes values $v = 1, \ldots, N$, with $v + N \equiv v$. The algebras generated by $\{Z_v, X_v X_{v + 1}\}$ and $\{Z_{v + \frac12}^\vee, X^\vee_{v - \frac12} X^\vee_{v + \frac12}\}$ are obviously isomorphic. Each algebra has a center generated by the product of all the $Z$ operators; this is a global kinematic $\Z_2$ symmetry. (A symmetry is kinematic if its generators form the center of the algebra in question, i.e.~if every Hamiltonian belonging to this algebra must have this symmetry.) Naively, this would mean that states that map under this duality are eigenstates of
\bel{
  Q \equiv \prod_v Z_v
}
(and their statistical mixtures) in the original theory, and eigenstates of
\bel{
  Q^\vee \equiv \prod_v Z^\vee_{v + \frac12}
}
(and their mixtures) in the dual theory; one might also expect that the global $\Z_2$ symmetries may just map to each other. This turns out to be wrong after studying the global consequences of the above local relations between generators.

The duality \eqref{1d KW original} actually produces a \emph{constraint}. By taking a product over all $v$ the following consistency relations emerge:
\bel{
  Q^\vee = \prod_v X_v X_{v + 1} = \1 \quad\trm{and}\quad Q = \prod_v X^\vee_{v - \frac12} X^\vee_{v + \frac12} = \1.
}
These conditions must be understood as promises that the only states that will ever be considered are those in which these operator equations are obeyed: density matrices of all other states do not belong to the algebras generated by $\{Z_v, X_v X_{v + 1}\}$ or $\{Z_{v + \frac12}^\vee, X_{v-\frac12}^\vee X_{v + \frac12}^\vee\}$. Thus the $\Z_2$ symmetry is projected to its singlet sector on both sides of the duality.\footnote{Projecting to the singlet sector is similar, but not the same as gauging the theory with a given symmetry. The former just means that a global constraint is imposed on the theory, and that only operators commuting with this constraints are allowed. On the other hand, a theory is \emph{gauged} if it is coupled to the appropriate gauge field and a local Gauss constraint is imposed. The global singlet constraint follows from multiplying all the local constraints. If a theory is \emph{topologically gauged}, the gauge field is taken to be topological, i.e.~the curvature of the gauge field is fixed. Topologically gauging is the closest one can get to projecting to the singlet sector while only imposing local constraints. Topologically gauged and singlet-projected theories only differ by holonomies on manifolds that are not simply connected. For examples where this difference is crucial, see \cite{Banerjee:2012gh, Banerjee:2013mca}.} Another way to say this is that ordinary KW is a map between two spin systems in which the modes with zero spatial momentum are not dynamical. This is a self-duality.\footnote{It is wrong to say that $\Z_2$ is spontaneously broken on both sides of the duality. In a pair of KW-dual theories every state in the spectrum must be a $\Z_2$-singlet, and there is generically no ground state degeneracy.}

Note that the standard KW duality cannot be implemented in every spin chain: the Hamiltonian must have a well-defined mapping under \eqref{1d KW original}. The Ising model \eqref{def Ising} can be dualized, but its deformation
\bel{\label{def Ising deformed}
  H\_{Ising}(h\_x) \equiv H\_{Ising} + h\_x \sum_v X_v
}
does not have a KW dual because it cannot be consistently projected to the singlet sector.

An alternative KW transformation can be applied to arbitrary spin systems, however. It can be obtained by adding topological gauge fields to one side of the original KW duality; appropriate local gauge constraints that mix matter and gauge fields are assumed, otherwise adding the gauge field would be a wholly trivial procedure. This procedure for generating dualities works in any dimension; in $d = 1$ gauge theories are automatically topological so here there is no need to be restricted to a particular class of Hamiltonians for the gauge sector. Consider adding gauge fields to the theory on the dual lattice, whose links are labeled by integers (so, for instance, the dual link $v$ connects dual vertices $v - \frac12$ and $v + \frac12$). The gauge field position and momentum operators will be denoted $Z_v^\vee$ and $X_v^\vee$. The Gauss operators of interest are
\bel{
  G^\vee_{v + \frac12} \equiv X^\vee_v Z^\vee_{v + \frac12} X^\vee_{v + 1}.
}
The gauge-invariant algebra consists of operators that commute with all the $G^\vee$'s. Its generators are electric fields $X_v^\vee$, matter position operators $Z_{v + \frac12}^\vee$, and ``covariant'' matter kinematic operators $X_{v - \frac12}^\vee Z_v^\vee X_{v + \frac12}^\vee$. The product of the latter gives the Wilson loop
\bel{
  W^\vee \equiv \prod_v Z^\vee_v.
}

Gauging means working only within the subspace in which $G^\vee_{v + \frac12} = \1$. This subspace is $2^N$-dimensional: it is a direct product of the $2^{N - 1}$-dimensional space of ``matter singlets'' from the previous passage and of the two-dimensional space of $W^\vee$-eigenstates. It is now possible to dualize the entire original spin chain, via
\algns{\label{1d KW alternative}
  Z_v &= X^\vee_{v - \frac12} Z^\vee_v X^\vee_{v + \frac12},\\
  X_v &= X^\vee_v.
}
The old mapping $X_v X_{v + 1} = Z_{v + \frac12}^\vee$ from \eqref{1d KW original} can be derived from this one by dualizing both $X^\vee$'s that appear in the Gauss law $G^\vee_{v + \frac12} = \1$. Note that the $\Z_2$ charge of the original system maps to the dual Wilson line, $Q = W^\vee$. Another interpretation of this Wilson line is the phase/monodromy associated to transporting an excitation along the spatial circle. (This intuition is precise in a gapped phase.) In other words, $Q$ dualizes to a boundary condition, which is in this case a fully quantum degree of freedom.

The duality \eqref{1d KW original} simply maps $h \mapsto 1/h$ in the Ising model \eqref{def Ising} (as two Hamiltonians can be considered equivalent if they differ by an overall rescaling).  The deformed Ising Hamiltonian \eqref{def Ising deformed} dualizes to
\bel{
  H\_{Ising}(h\_x) = \sum_v \left(Z^\vee_{v - \frac12} + h X^\vee_{v - \frac12} Z_v^\vee X^\vee_{v + \frac12} + h\_x X^\vee_v\right).
}
Now the gauge fields enter the Hamiltonian, and $1/h$ can be interpreted as the transverse field while $h\_x/h$ can be interpreted as the gauge coupling.

\subsection{Jordan-Wigner} \label{subsec 1d JW}

Jordan-Wigner (JW) duality maps the original Ising spin theory to that of two Majorana operators per site,
\algns{\label{1d JW original}
  \chi_v &= Z_1 Z_2 \cdots Z_{v - 1} X_v,\\
  \chi_v' &= Z_1 Z_2 \cdots Z_{v - 1} Y_v.
}
A beginning point must be chosen, but its choice does not affect the physics. The paramagnetic state with $Z_v = \1$ on all sites maps to the state with no fermions, denoted $\qvec 0$. The rest of the Hilbert space maps follow from this convention and eq.~\eqref{1d JW original}, by acting on $\qvec 0$ with different operators.

It is convenient to define bosonic operators that are fermion bilinears,
\bel{\label{def S Z}
  S_{vu} \equiv - \i \chi'_v \chi_u, \quad Z_v \equiv \i \chi'_v \chi_v.
}
The $Z_v$ built out of fermions is the same as the bosonic one (hence the same label), and it measures the fermion number at site $v$. Meanwhile, $S_{v, v+1}$ moves an excitation between $v$ and $v + 1$, and its bosonic dual is
\algns{
  S_{v, v+ 1} &= X_v X_{v + 1}\quad \trm{for}\ v = 1, \ldots, N - 1,\\
  S_{N, 1} &= Y_1 Z_2 \ldots Z_{N - 1} Y_N = - Q X_1 X_N.
}
This means that
\bel{
  \prod_v S_{v, v + 1} = - Q,
}
so the transport of a fermion around the circle depends on how many other fermions there are along the way.\footnote{This is intuitive: if there is an even number of fermions in a state ($Q = \1$), transporting one fermion means commuting it past an odd number of fermions. Hence the minus sign. It is important to stress that the fermionic excitations are transported by $S_{vu}$, not by $\psi\+_v \psi_u$, so in this picture they are not ``hard-core'' and can pass through each other while leaving a minus sign in the wavefunction.} Note that here there is no need to introduce a gauge field to get nontrivial global properties of fermion transport, unlike in the KW story. Indeed, JW duality maps the entire $2^N$-dimensional space of Ising spins to the fermionic system from the get-go.

The gauged spin system on the dual lattice in the map \eqref{1d KW alternative} can be replaced with $\Z_2$ QED\footnote{By the conventions of Kogut and Susskind \cite{Kogut:1974ag}, the lattice Hamiltonian for Maxwell theories with $\Z_2$ gauge group and coupling $g$ is $H\_{gauge} = g^2 \sum_v X_v^\vee$ in $d = 1$. ``Zero coupling'' means $g = 0$, implying $H\_{gauge} = 0$. The matter still has unit charge under the gauge symmetry, which means that it is coupled to the gauge fields via the Gauss operators $G^\vee_{v + \frac12}$ in \eqref{1d gauge constraint}.} (with spinless fermions) by applying a JW transformation on the matter variables $X_{v +\frac12}^\vee$ and $Z_{v + \frac12}^\vee$. This gives the following map between an ordinary spin chain and a theory of spinless fermions on dual sites coupled to $\Z_2$ gauge fields:
\algns{\label{1d JW alternative}
  Z_v &=  S_{v - \frac12,\, v+\frac12}^\vee Z^\vee_v  \equiv -\i {\chi'}^\vee_{\!v - \frac12} Z_v^\vee \chi^\vee_{v + \frac12} \quad \trm{for}\quad v = 1, \ldots, N - 1,\\
  Z_N &= - S^\vee_{N - \frac12, \frac12} Z_N^\vee \equiv \i {\chi'}^\vee_{\!N- \frac12} Z_N^\vee \chi^\vee_{\frac12},\\
  X_v &= X_v^\vee,
}
with the gauge constraint
\bel{\label{1d gauge constraint}
  G^\vee_{v + \frac12} = X_v^\vee Z_{v + \frac12}^\vee X_{v + 1}^\vee = \1
}
as before. These relations imply
\bel{
  Q = W^\vee \quad \trm{and} \quad Q^\vee \equiv \prod_v \i {\chi'}_{v + \frac12}^\vee \chi^\vee_{v + \frac12} = \1,
}
consistent with the previous dualities. The different mapping of $Z_N$ in \eqref{1d JW alternative} is crucial for overall consistency. In particular, the Ising model \eqref{def Ising} maps to a $\Z_2$ QED at zero coupling with antiperiodic boundary conditions (BCs) for fermions.

The word ``antiperiodic'' is subtle: the algebra does not know about periodicity and there is no fixed monodromy due to a gauge field. However, by \eqref{1d JW alternative} the term $\sum_{v = 1}^N Z_v$ in $H\_{Ising}$ dualizes to $\sum_{v = 1}^N (-\i) {\chi'}^\vee_{v - \frac12} Z_v^\vee \chi^\vee_{v + \frac12}$ with the convention $\chi^\vee_{v + \frac12 + N} = - \chi^\vee_{v + \frac12}$.\footnote{The Hamiltonian does not feature ${\chi'}_{v + \frac12 + N}^\vee$, so there is no need to define a specific boundary condition for this operator. A different Hamiltonian could have lead to a natural boundary condition for $\chi'$, too, and in that case the entire complex fermion $\psi$ would have been given a natural boundary condition.} Antiperiodicity thus only enters at the level of the Hamiltonian. For instance, for Hamiltonians which do not feature terms like $\sum_v Z_v$, such as e.g.~$\sum_v X_v$, there is no reason to say that the fermions are antiperiodic. When applying the ordinary JW \eqref{1d JW original} to the Ising model \eqref{def Ising}, $\sum_v X_v X_{v + 1}$ maps to $S_{12} + S_{23} + \ldots + S_{N - 1, N} - Q S_{N, 1}$, and the states with $Q = \1$ are said to have antiperiodic (Neveu-Schwarz) BCs while states with $Q = -\1$ have periodic (Ramond) ones.\footnote{Boundary conditions that depend on a global charge like $Q$ will be called \emph{dynamical}. In any $d$, dynamical boundary conditions will depend on one-form charges. In $d = 1$, one-form charges are the same as familiar zero-form charges.} The upshot of this discussion is that the notion of periodicity for fermions is \emph{context-dependent} and only has meaning relative to a specific class of Hamiltonians; in the above example, the notion of boundary conditions arose naturally from the fact that $H\_{Ising}$ was a sum of $N - 1$ terms of form $S_{v, v+1}$ and one term $-QS_{N,1}$. More formally, this context-dependence reflects the fact that the space of spin structures is an \emph{affine} space, as will be discussed in some more detail in subsection \ref{subsec 2d spin structures}.

\subsection{Twisted dualities} \label{subsec twisting}

The dualities discussed so far, applied to Ising spins in $d = 1$, are
\boxedAlgns{
  \trm{spins}/\Z_2 &= \trm{spins}^\vee/\Z_2, \\
  \trm{spins} &= \trm{gauged\ spins}^\vee \\
  &= \trm{fermions\ with\ dynamical\ BCs} \\
  &= \Z_2\trm{\ QED}^\vee\trm{\ with\ antiperiodic\ BCs}.
}
Recall that checks (${}^\vee$) indicate that the theory lives on a dual lattice, and ``$/\Z_2$'' denotes projection to the singlet sector of the global $\Z_2$ symmetry. The notion of (anti)periodicity on the fermion side is the one inherited from the Ising model \eqref{def Ising} as discussed in the previous subsection. Different Hamiltonians may yield different boundary conditions.

There are more dualities that can be obtained from the ones above. In particular, dualities that map singlet sectors of global symmetries can be modified so that the singlet sector of one symmetry is mapped to a fixed nonzero charge sector of another symmetry. These modifications of dualities will be called \emph{twists}. One way to think about them is as classical background fields that must be turned on for a particular duality to hold.

Consider the following twist of KW duality \eqref{1d KW original} for $\eta_v^\vee \in \{\pm 1\}$:
\algns{\label{1d KW twisted}
  Z_v &= \eta_v^\vee X^\vee_{v - \frac12} X^\vee_{v + \frac 12} \\
  X_v X_{v + 1} &= Z^\vee_{v + \frac12}.
}
The duality is consistent if
\bel{
  Q = \prod_{v = 1}^N \eta_v^\vee \quad \trm{and} \quad Q^\vee = \1.
}
Thus, choosing e.g.~$\eta_N^\vee = -1$ and $\eta_1^\vee = \ldots = \eta^\vee_{N - 1} = 1$ gives a duality between the $Q = -\1$ sector of the original theory and the singlet, $Q^\vee = \1$, sector of the dual theory. Introducing twisting variables $\eta_{v + 1/2}$ in the second line of \eqref{1d KW twisted} can further change the duality to be between the $-\1$ sectors of each theory. The singlet sector constraints do not change under ``gauge transformations'' that flip the sign of two $\eta_v^\vee$'s or $\eta_{v + \frac12}$'s.

The second KW duality \eqref{1d KW alternative} now follows from promoting the twists $\eta_v^\vee$ to dynamical variables $Z^\vee_v$. The operators that change $Q$ on the original lattice can be mapped to operators that change $\prod_v \eta_v^\vee$ on the dual lattice, namely $X_v^\vee$. In general, all that is needed is to promote $\prod_v \eta_v^\vee$ into a new $\Z_2$ degree of freedom, but it is more natural to do this locally by replacing each $\eta_v^\vee$ with $Z_v^\vee$ and then imposing gauge constraints at dual sites.
\bigskip

The JW map \eqref{1d JW original} is not a ``singlet-singlet'' duality, so there is no reason to twist it. However, consider the following mapping of bilinears $S_{vu}$ and $Z_v$:
\algns{\label{1d JW restricted}
  Z_v &= \i \chi_v' \chi_v,\\
  X_v X_{v + 1} &=  -\i \chi'_v \chi_{v + 1} \equiv S_{v, v+1}.
}
This is a new singlet-singlet duality: the second line (which holds for all $v = 1, \ldots, N$ and assumes that $\chi_{N + 1} = \chi_1$) implies that \eqref{1d JW restricted} maps the $Q = -\1$ sectors to each other. (Recall that the original duality \eqref{1d JW original} had $X_v X_{v + 1} = (-Q)^{\delta_{v,N}} S_{v, v+ 1}$.) The map \eqref{1d JW restricted} can first be twisted to give
\algns{\label{1d JW restricted twisted}
  Z_v &= \i \chi_v' \chi_v,\\
  X_v X_{v + 1} &= \eta_{v + \frac12} S_{v, v+1},
}
with the last line again applying to all $v = 1,\ldots, N$. Choosing the c-numbers $\eta_{v + \frac12}$ such that their product is $-1$ gives another singlet-singlet duality, this time with $Q = \1$ on both sides.

Now consider again promoting $\eta_{v + \frac12}$ to a $\Z_2$ gauge field $Z_{v + \frac12}$, getting a tentative duality of the form%
\algns{\label{trial JW duality}
  Z_v &= \i \chi_v' \chi_v,\\
  X_v X_{v + 1} &= -\i \chi_v' Z_{v + \frac12} \chi_{v + 1} = S_{v, v+1} Z_{v + \frac12}.
}
The second line implies a global consistency condition
\bel{
  QW = - \1
}
with $W \equiv \prod_v Z_{v + \frac12}$. (The dual of $\sum_v X_v X_{v + 1}$ in this case is a fermion with periodic BCs.) If the goal is to get a full duality involving all sectors of $Q$, the theory of fermions and gauge fields cannot be $\Z_2$ QED like in \eqref{1d JW alternative}, i.e.~the gauge constraint cannot be $X_{v - \frac12} Z_v X_{v + \frac12} = \1$. Such a constraint would imply $Q = \1$ and hence $W = -\1$, completely freezing out all $\Z_2$ gauge degrees of freedom and giving back the singlet-singlet duality \eqref{1d JW restricted twisted} with $\prod_v \eta_{v + \frac12} = -1$.

Consider, however, the following unusual gauge constraint:
\bel{\label{1d flux attached gauge constraint}
  X_{v - \frac12} Z_v X_{v + \frac12} = (-W)^{\delta_{v, 1}}.
}
Both $Z_v$ and $S_{v, v + 1} Z_{v + \frac12}$ are gauge-invariant under this constraint, and taking a product over all $v$ gives $QW = -\1$ as required. It is easy to check that the full duality is
\algns{\label{1d JW original 2}
  Z_v &= \i \chi_v' \chi_v,\\
  X_v &= X_{\frac12} (\i \chi_1' \chi_1) Z_{1 + \frac12} (\i \chi_2' \chi_2) \cdots Z_{v - \frac12} \chi_v,
}
with the operators on the r.h.s.~being the generators of the gauge-invariant algebra appropriate to the modified constraint \eqref{1d flux attached gauge constraint}. The ordinary JW transformation follows after fully gauge-fixing because the Wilson line $W$ is constrained to equal $-Q$.

Note the difference between twisting \eqref{1d KW original} and \eqref{1d JW restricted}: in the former case, the Wilson loop was necessary to achieve the full mapping \eqref{1d KW alternative}, while in the latter case the Wilson loop ended up being constrained and tradeable for a matter degree of freedom in the full mapping \eqref{1d JW original}. This last phenomenon is a $d = 1$ coincidence: generically the topological degrees of freedom will be there to stay after twisting.

\subsection{A comment on orientations}

The above analysis tacitly assumed that all the links were directed the same way, say from $v$ to $v + 1$. This choice was reflected in working with $S_{uv} = -\i \chi'_u \chi_v$ as opposed to $S'_{uv} = -\i \chi'_v \chi_u$: the algebra of fermion bilinears is generated by one fermion number operator $Z$ per site and by one hopping operator $S$ per link, and it was a matter of convention to always pick $S_{v, v+ 1}$ as the generator associated to link $v + \frac12$. Changing this convention, or equivalently changing the orientation of a link, affects the form the duality takes. Indeed, substituting $S_{v, v + 1} \mapsto S'_{v, v + 1}$ in \eqref{1d JW restricted} does not lead to a duality, and no choice of twisting changes this fact. To recover a duality, the other side of the mapping must be changed as well, getting e.g.~$Y_v Y_{v + 1} = S'_{v, v+ 1}$. In general, there is thus no well-defined action of orientation change on the periodicity of fermions (i.e.\ on the twists $\eta_{v + \frac12}$). A very special exception to this will appear in subsection \ref{subsec 2d spin structures}, where a particular class of orientations will be identified with gauge field variables in a 2D path integral.

%\textsf{\small \textbf{Internal note:} In $d = 2$, orientations satisfying the Kasteleyn property are known to induce spin structures in a canonical way. (A perfect matching of vertices is needed to make this canonical; this is assumed to be fixed.) It thus seems that each Kasteleyn orientation corresponds to a particular boundary condition for fermions. This doesn't contradict the previous passage because boundary conditions depend on the underlying Hamiltonian. The association of spin structures to boundary conditions in the continuum is defined relative to the Dirac Hamiltonian, $H = \i \bar \Psi \c \del \Psi$, while on the lattice this is done relative to e.g.~Cimasoni's discrete Dirac operator. Here the orientation change does induce a change in boundary conditions.}

\section{$\Z_2$ dualities in $d = 2$}

\begin{figure}[tb!]
\begin{center}
\begin{tikzpicture}[scale = 2]

\draw[step = 0.5, dotted, thick] (-2, -1) grid (2, 1);
\draw[red, step = 0.5, xshift = 0.25cm, yshift = 0.25cm, dotted] (-2.5, -1.5) grid (2, 1);
\draw[white, very thick] (-2.25, -1.25) rectangle (2.25, 1.25);

\begin{scope}[thick, decoration={
    markings,
    mark=at position 0.5 with {\arrow{>}}}
    ] 
    \draw[postaction={decorate}] (-1.5, -0.5) -- (-1.5, 0);
\end{scope}
\draw (-1.55, -0.25) node[anchor = east] {\contour{white}{$\ell$}};
\draw[fill = black] (-1.5, -0.5) circle (0.025);
\draw[fill = black] (-1.5, 0) circle (0.025);
\draw (-1.5, -0.5) node[anchor = north] {\contour{white}{$\ell_1$}};
\draw (-1.5, 0) node[anchor = south] {\contour{white}{$\ell_2$}};

\draw[fill = black] (0, 0.5) circle (0.025);
\draw[red, thick] (-0.25, 0.25) rectangle (0.25, 0.75);
\draw (0, 0.5) node[anchor = north] {{\contour{white}{$v \in \Mbb$}}};
\draw (0.25, 0.75) node[anchor = west, red] {{\contour{white}{$v \in \Mbb^\vee$}}};

\draw[thick] (1, -0.5) rectangle (1.5, 0);
\draw[red, fill = red] (1.25, -0.25) circle (0.025);
\draw (1, -0.5) node[anchor = north] {{\contour{white}{$f \in \Mbb$}}};
\draw (1.25, -0.25) node[anchor = west, red] {{\contour{white}{$f \in \Mbb^\vee$}}};

\end{tikzpicture}
\end{center}
\caption{\small A rectangular lattice $\Mbb$ (black) and its dual lattice $\Mbb^\vee$ (red). The figure depicts examples of an oriented link $\ell$ and its endpoints $\ell_{1/2}$, a site $v \in \Mbb$ and its dual face, and a face $f$ and its dual site. In this section, $\Z_2$ degrees of freedom can be associated either to vertices, edges, or faces of both $\Mbb$ and $\Mbb^\vee$.
}
\label{fig 2d lattices}
\end{figure}
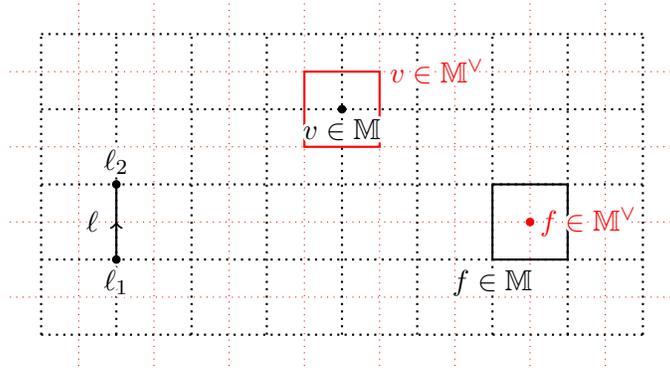

Consider a lattice $\Mbb$ with oriented links $\ell$ (fig.~\ref{fig 2d lattices}). The two vertices belonging to $\ell$ will be denoted $\ell_1$ and $\ell_2$, with the convention that the link is oriented from $\ell_1$ towards $\ell_2$. Faces (plaquettes) $f \in \Mbb$ also admit a canonical enumeration $f_0, f_1, \ldots$ of their constituent vertices if the orientation of links in $\Mbb$ forms a \emph{branching structure}, which will be discussed more below. The dual lattice $\Mbb^\vee$ has faces labeled by $v$, links labeled by $\ell$, and vertices labeled by $f$, and as before operators that live on $\Mbb^\vee$ will be denoted by a check ($^\vee$). Abusing the notation a bit, $\Mbb$ and $\Mbb^\vee$  will henceforth also denote the spaces of $k$-chains $C_k(\Mbb, \Z_2)$ and $C_k(\Mbb^\vee, \Z_2)$ for each $k = 0, 1, 2$. In section \ref{sec Zk} each appearance of $\Z_2$ should be replaced with $\Z_K$.

Recall a few more terms that will be used later in this paper. A boundary operator $\del$ acts on chains in the usual way, lowering their order. For instance, the boundary of a link $\ell$ is the zero-chain $\del\ell = \ell_1 + \ell_2$. Boundary operators on the dual lattice will be denoted $\del^\vee$. A chain with no boundary is called a cycle. A $k$-cochain is a function from $k$-chains (e.g.~sites, links, faces) to $\pm 1$, with the composition of cochains obtained by multiplication.\footnote{Note the further notation abuse for the sake of convenience: chains are composed by addition, with coefficients in $\{0, 1\}$, while cochains are multiplied. ``Cochains'' in this paper are really exponentials of cochains that are composed additively.} Coboundary operators $\delta$ increase the order of cochains and are analogous to exterior derivatives of forms on manifolds. For instance, the coboundary of a ``delta function'' on site $v$ --- a zero-cochain $\Delta^v$ that is $-1$ on $v$ and $+1$ elsewhere --- is a product of delta functions on links containing $v$, $\delta\Delta^v = \prod_{\ell \supset v} \Delta^\ell$.

Subsections \ref{subsec 2d KW} and \ref{subsec 2d JW} generalize the KW and JW dualities from the previous section. In particular, subsection \ref{subsec 2d JW} introduces a number of important ideas, defining flux attachment and demonstrating the anomaly of the one-form symmetry in whose singlet sector a dual of a fermionic theory must be. Subsection \ref{subsec 2d spin structures} reviews some background on spin structures and shows how it is reasonable to identify them with twists of JW dualities.

\subsection{Kramers-Wannier and its twisting} \label{subsec 2d KW}

The standard $\Z_2$ version of KW duality applies to spins living on sites in $\Mbb$. It is given by
\algns{\label{2d KW original}
  X_{\ell_1} X_{\ell_2} &= X_\ell^\vee, \\
  Z_v &= W^\vee_v.
}
The operators on the right-hand side are local generators of the gauge-invariant algebra of a pure $\Z_2$ gauge theory.\footnote{The \emph{full} algebra of this gauge theory is generated by a pair of Pauli matrices $X_\ell^\vee, Z^\vee_\ell$ on each link. This algebra is not gauge-invariant, but it is a useful concept to define \cite{Lin:2018bud}. Gauge-invariant operators are those that commute with Gauss operators, $G_f^\vee \equiv \prod_{\ell \subset f} X_\ell^\vee$, on all sites $v$. The algebra of gauge-invariant operators is generated by $X_\ell^\vee$, $W_v^\vee \equiv \prod_{v\supset \ell} Z_\ell^\vee$, and by nonlocal operators $W^\vee_c \equiv \prod_{\ell \subset c} Z^\vee_\ell$ along noncontractible one-cycles on $\Mbb$.} They are, respectively, the electric fields on dual links and Wilson loops along dual faces.

Consistency again enforces a singlet constraint on both theories. The second line of \eqref{2d KW original} enforces
\bel{
  Q \equiv \prod_v Z_v = \1,
}
like before. The first line enforces a new kind of constraint,
\bel{\label{def T}
  T_c^\vee \equiv \prod_{\ell \subset c} X_\ell^\vee = \1,
}
where $c$ is a one-cycle on the \emph{original} lattice $\Mbb$. If $c$ is contractible, this constraint is equivalent to the Gauss law in the interior of $c$ on $\Mbb^\vee$. In particular, if $c$ is just a plaquette $f$, the constraint is $G^\vee_f = \1$. On the other hand, if $c$ is noncontractible, this is a genuinely new constraint. The operators $T_c^\vee$ in a gauge theory generate one-form symmetries, so the gauge theory can be understood to be projected to the singlet sector of a one-form symmetry.

Both singlet constraints can be twisted. The analogue of the $d = 1$ twist \eqref{1d KW twisted} is
\algns{\label{2d KW twisted}
  \eta_\ell X_{\ell_1} X_{\ell_2} &= X_\ell^\vee, \\
  Z_v &= W^\vee_v.
}
This affects the one-form constraint, which becomes
\bel{
  T^\vee_c = \prod_{\ell \subset c} \eta_\ell.
}
Note that ``gauge transformations'' of the $\eta_\ell$'s do not change the singlet constraint: for any $v$ and $c$, the product $\prod_{\ell \subset c} \eta_\ell$ stays the same when all twists $\eta_\ell$ for $\ell \supset v$ are flipped. Indeed, like before, the twists can be replaced with topological gauge fields, leading to the duality
\algns{\label{2d KW alternative}
  X_{\ell_1} Z_\ell X_{\ell_2} &= X_\ell^\vee, \\
  Z_v &= W^\vee_v, \\
  W_c = T^\vee_c, &\quad  T_{c^\vee} = W_{c^\vee}^\vee,
}
where $c^\vee$ is a noncontractible one-cycle on $\Mbb^\vee$. The various constraints implicit in the theories above are
\bel{
  W_f = \1, \quad G_v \equiv Z_v \prod_{\ell \supset v} X_\ell = \1, \quad G_f^\vee = \1.
}
Note that $G_v$ takes matter into account, and hence it is not the same as the Gauss operator in a pure gauge theory. The first requirement --- that all gauge fields be topological/flat in the gauged matter theory --- means that all electric fields $X_\ell$ have zero expectations except when they are multiplied to form the operators $T_c$. The second constraint, multiplied over all $v$, ensures that $Q = \1$.
\bigskip

Twisting the $Q = \1$ singlet constraint in eq.\ \eqref{2d KW original} gives
\algns{
  X_{\ell_1} X_{\ell_2} &= X_\ell^\vee,\\
  Z_v &= \eta_v^\vee W_v^\vee,
}
and the singlet constraint in the matter theory is now $Q = \prod_v \eta^\vee_v$. Making the twists dynamical corresponds to coupling a topological one-form gauge theory to the ordinary $\Z_2$ gauge theory.\footnote{The full algebra of a one-form $\Z_2$ gauge theory is generated by $Z_v^\vee$ and $X_v^\vee$ for all dual plaquettes $v$. The Gauss operators in a pure one-form theory are $G^\vee_\ell = \prod_{v \subset \ell} X_v^\vee$, which means that the gauge-invariant algebra is generated by $X_v^\vee$ and $W_{\Mbb}^\vee = \prod_v Z_v^\vee$ for a connected spatial manifold/lattice. In $d = 2$, every one-form gauge theory is topological.} This is done by charging the electric fields of the ordinary gauge theory, i.e.~by adding the one-form degrees of freedom and then by imposing the one-form Gauss constraint
\bel{\label{2d one-form gauge constraint}
  G_\ell^\vee \equiv X_{\ell_1}^\vee X^\vee_\ell X_{\ell_2}^\vee = \1.
}
(Recall that $\ell_{1/2}$ are vertices at the ends of link $\ell$, so $X_{\ell_{1/2}}^\vee$ are operators on dual faces that share dual link $\ell$.) The resulting duality is then
\algns{\label{2d KW alternative 2}
  X_v &= X^\vee_v,\\
  Z_v &= Z_v^\vee W_v^\vee,
}
and the operators on the r.h.s.~are precisely the one-form-gauge-invariant operators, i.e.~the operators that commute with the Gauss operators $G_\ell^\vee$ in \eqref{2d one-form gauge constraint}. The theory on the r.h.s.~will be called --- perhaps unfortunately --- a \emph{gauged $\Z_2$ gauge theory}.

The only constraints implicit in \eqref{2d KW alternative 2} are the one-form Gauss laws \eqref{2d one-form gauge constraint}. Multiplying these over dual links $\ell \subset c$ gives back \eqref{def T}, namely
\bel{
  T^\vee_c = \1.
}
When $c$ is a contractible cycle, say the boundary of a face ($c = \del f$), this becomes the Gauss law, $G_f^\vee = \1$. Thus the zero-form Gauss law is \emph{contained} in the one-form Gauss law. When $c$ is not contractible this is a constraint that makes sure that the holonomies $W^\vee_c$ all have vanishing expectations. Note that another map that follows from \eqref{2d KW alternative 2} is
\bel{
  Q = W_{\Mbb}^\vee \equiv \prod_v Z_v^\vee,
}
the relation between the ordinary symmetry generator of spin systems and the Wilson ``surface'' operator of the one-form gauge theory.
\bigskip

To recap, KW dualities involving Ising spins in $d = 2$ studied here are
\boxedAlgns{
  \trm{spins}/\Z_2 &= \Z_2\ \trm{gauge\ theory}^\vee/(\Z_2\ \textrm{one-form}),\\
  \trm{topologically\ gauged\ spins} &= \Z_2\ \trm{gauge\ theory}^\vee, \\
  \trm{spins} &= \trm{gauged\ }\Z_2\ \trm{gauge\ theory}^\vee.
}
As before, more dualities can be obtained from these. For instance, the second line above can be modified into a self-duality of two gauged spin systems, both with dynamical (not necessarily topological) gauge fields.

\subsection{Jordan-Wigner and its twisting} \label{subsec 2d JW}

The higher-dimensional analogues of JW duality have been formulated as exact dualities only recently \cite{Chen:2017fvr, Chen:2018nog} (see \cite{Chen:2017fvr} for references on previous proposals on higher-dimensional bosonization). The key insight is that the $\Z_2$ gauge theory dual to a fermion theory has a nonstandard Gauss law. This law implements flux attachment, i.e.~it ensures that a magnetic flux is also electrically charged. A similar constraint was encountered in eq.~\eqref{1d flux attached gauge constraint}, where changing a magnetic flux through the disk enclosed by the spatial circle also changed the electric charge at one of the sites on the circle.

In the standard form of the $d = 2$ JW duality, as presented in \cite{Chen:2017fvr}, pairs of Majoranas live on sites in $\Mbb$ and their bilinears map as
\algns{\label{2d JW twisted}
  \eta_\ell S_\ell &= \~X_\ell^\vee,\\
  Z_v &= W^\vee_v.
}
Here $S_\ell = -\i \chi'_{\ell_1} \chi_{\ell_2}$ is the hopping generator associated to $\ell$ via a fixed orientation of links on $\Mbb$. The dual lattice has $\Z_2$ variables living on its links, but the corresponding Gauss operators are
\bel{\label{2d flux attached gauge constraint}
  \~G_f^\vee \equiv G_f^\vee \prod_{v:\, v_0 = f} W_v^\vee.
}
The product above goes over all dual faces $v$ whose anchor $v_0$ is precisely the dual vertex $f$.\footnote{Each dual face is assumed to have one of its vertices chosen as an anchor. Just like vertices belonging to a link $\ell$ are labeled $\ell_{1/2}$ depending on the orientation, vertices of a face $f$ can be canonically labeled $f_{0/1/2/\ldots}$ if the orientations on links form a branching structure; then the vertex $f_0$ is called the anchor of face $f$. Conversely, the anchor of a dual face $v$ is a dual vertex denoted $v_0$. The orientation of links on $\Mbb$ induces at least one consistent assignment of anchors to dual faces on $\Mbb^\vee$, as discussed in the main text. If $\Mbb$ satisfies some additional properties, there is an algorithmic way to find the corresponding branching structure (and hence anchor assignment) on $\Mbb^\vee$ \cite{Wang:2017moj}.} The Gauss law thus also causes a magnetic flux through a dual face $v$ to induce an electric charge at the dual vertex $v_0$. Gauge-invariant operators must commute with $\~G_f^\vee$ (but not necessarily with $G_f^\vee = \prod_{\ell \subset f} X_\ell^\vee$), and the algebra of gauge-invariant operators is generated by Wilson loops $W_v^\vee$ on dual faces, by Wilson loops $W^\vee_c$ on noncontractible one-cycles $c$, and by modified electric field operators
\bel{\label{def tilde X}
  \~X^\vee_\ell \equiv X^\vee_\ell \prod_{\ell' \subset c(\ell)} Z_{\ell'}^\vee.
}
This product runs over all the links in the one-chain $c(\ell) \subset \Mbb^\vee$ that connects the anchors of dual faces on the two sides of $\ell$. Electric field operators thus must come accompanied with a Wilson line on some of the adjacent links. This way operators $\~X^\vee_\ell$ on different links $\ell$ can still fail to commute. For a fixed orientation of links on $\Mbb$, the anchors and the chains $c(\ell)$ can be chosen on $\Mbb^\vee$ so $\~X_\ell^\vee$ are gauge-invariant and have the same commutation relations as $S_\ell$.\footnote{\label{foot sq lattice}This choice is not necessarily unique. Consider a square lattice $\Mbb$ with all horizontal links oriented eastward, and all vertical links oriented northward. Then there are two choices \emph{per face} for consistent flux attachment: the anchor of every dual face can be either in its northeast or in its southwest corner.} Physically, $\~X_\ell^\vee$ transports a magnetic flux together with its attached electric charge across the dual link $\ell$.

The duality \eqref{2d JW twisted} already comes with twists $\eta_\ell$ included because on a generic lattice it is not possible to maintain consistency while setting them all to unity. Put another way, depending on the lattice, the fermions may need be coupled to a background gauge field with nontrivial curvature in order to get a consistent duality. To illustrate this better, consider a simple (not self-intersecting), possibly contractible one-cycle $c \in \Mbb$. Traversing $c$ in one direction (starting from an arbitrary site), the path-ordered product of hopping operators obeys
\bel{\label{2d proto vertex relation}
  \prod_{\ell \subset c} S_\ell = - \sideset{}{'}\prod_{v \subset c} Z_v,
}
where the primed product runs over all the vertices in $c$ that connect links with the same orientation relative to the direction of traversal of $c$. If $c$ encircles a single face $f$, i.e.~if $c = \del f$, then \eqref{2d proto vertex relation} is a \emph{vertex relation} at dual vertex $f$:
\bel{\label{2d vertex relation}
  \prod_{\ell \subset f} \left(\eta_\ell \~X^\vee_\ell \right)  = - \sideset{}' \prod_{v \subset f} W_v^\vee.
}
Thus
\bel{
  (\delta \eta)_f \equiv \prod_{\ell \subset f} \eta_\ell
}
must be chosen so that, for a fixed $c(\ell)$ in \eqref{def tilde X}, the above product agrees in sign with the modified Gauss law $\~G^\vee_f = \1$ based on eq.\ \eqref{2d flux attached gauge constraint}. Alternatively, a consistent duality can be obtained with $\eta_\ell = 1$ everywhere but with a nontrivial background charge density in the dual theory. As in the KW case, ``gauge transformations'' of $\eta_\ell$ do not affect the consistency of the duality.

When $c$ is not contractible, eq.~\eqref{2d proto vertex relation} is a constraint in the dual theory that can be written as\footnote{Another way to write this is
$$ (-1)^{\varphi(c)} T^\vee_c W^\vee_{\~c} = \1,$$
with $\varphi(c) \in \Z_2$ determined by the $\eta$'s and by orientations of links near and on $c$. The cycle $\~c \subset \Mbb^\vee$ is given by
$$\~c \equiv \sum_{\ell \subset c} c(\ell) + \sideset{}' \sum_{v \subset c} \del^\vee v,$$
where $\del^\vee$ is the boundary operator on chains in $\Mbb^\vee$, so the dual one-cycles $\del^\vee v$ are boundaries of dual faces $v$ that enter the product in eq.\ \eqref{2d vertex relation}. The distinction between $c$ and $\~c$ is one lattice analogue of the \emph{framing} necessary to define line operators with spin in continuum theories.}
\bel{
  \1 = -\prod_{\ell \subset c} S_\ell \sideset{}' \prod_{v \subset c} Z_v = -\prod_{\ell \subset c} \left(\eta_\ell \~X_\ell^\vee\right) \sideset{}' \prod_{v \subset c} W^\vee_v  \equiv \bigg(\prod_{\ell \subset c} \eta_\ell \bigg)\~T^\vee_c,
}
In this notation, the singlet-singlet nature of duality \eqref{2d JW twisted} is clear: the constraints involved are
\bel{
  Q = \prod_v Z_v = \1, \quad \~G_f^\vee = \1, \quad\trm{and}\quad \~T_c^\vee = \bigg(\prod_{\ell \subset c} \eta_\ell \bigg) \1
}
for a noncontractible $c$. In particular a ``standard'' JW duality could be defined as the one where $\~T^\vee_c = \1$, in analogy to how the ``standard'' KW has $T^\vee_c = \1$. Note that $\~T^\vee_c$ transports one electric and one magnetic excitation along a cycle, and its path-ordering makes sure it does not change under deformations of $c$ by contractible cycles. In toric code parlance \cite{Kitaev:1997wr}, it is a particular realization of the operator that moves $\eps$ excitations along $c$.
\bigskip

The twists $\eta_\ell$ can now be promoted to quantum variables, as before. The new duality is
\algns{\label{2d JW alternative}
  S_\ell Z_\ell &= \~X^\vee_\ell,\\
  Z_v &= W_v^\vee,\\
  W_c = \~T^\vee_c, &\quad   T_{c^\vee} = W_{c^\vee}^\vee.
}
Once again the gauge field on the fermion side must be placed in definite eigenstates of local Wilson loops $W_f$, but now the rule must be
\bel{
  W_f = (\delta \eta)_f,
}
which means that the duality can only be consistent on a generic lattice if the gauge fields are allowed to have the same nontrivial curvature that the background fields $\eta_\ell$ used to have. Other required constraints are usual,
\bel{
  G_v \equiv Z_v \prod_{\ell \supset v} X_\ell = \1, \quad \~G^\vee_f = \1.
}
The theory on the l.h.s.~of \eqref{2d JW alternative} is simply the theory of free fermions coupled to a topological gauge field. For particular classes of Hamiltonians, it can be interpreted as a theory in which all possible boundary conditions of fermions are summed over.
\bigskip

The other singlet constraint in \eqref{2d JW twisted} can be twisted, too. The twisted duality only differs by the mapping of $Z_v$:
\algns{\label{2d JW twisted 2}
  \eta_\ell S_\ell &= \~X_\ell^\vee,\\
  Z_v &= \eta^\vee_v W^\vee_v.
}
The vertex relation \eqref{2d vertex relation} now becomes
\bel{
  \prod_{\ell \subset f} \left(\eta_\ell \~X_\ell^\vee \right) = - \sideset{}' \prod_{v \subset f} \left(\eta^\vee_v W_v^\vee\right).
}
In most cases it is possible to set $\eta_\ell = 1$ on all links, as $\prod'_{v \subset f} \eta^\vee_v$ can be used to fix the sign here. However, this is not guaranteed to work on every lattice: it may happen that there are no vertices on $\del f$ where the orientation of links does not change. Hence it is wiser to keep $\eta_\ell$ fixed at some value that makes $\eta^\vee_v = 1$ a consistent choice. It will be assumed that $\prod_{\ell \subset c} \eta_\ell = 1$ for a noncontractible cycle. With this convention in place, the above relation can also be written as
\bel{\label{2d flux attached twisted gauge constraint}
  \~G^\vee_f = \sideset{}' \prod_{v\subset f} \eta^\vee_v.
}
The twisting also affects the line operator $\~T^\vee_c$, changing the one-form symmetry singlet condition to
\bel{
  \~T^\vee_c = \bigg(\sideset{}' \prod_{v \subset c} \eta^\vee_v\bigg) \1.
}
Note that the singlet conditions all stay the same under ``gauge transformations'' that change the signs of $\eta_v^\vee$ on two dual faces that share a dual link.

Unlike all previous examples, it is \emph{impossible} to simply promote this twisting by $\eta^\vee_v$ into a coupling between the flux-attached $\Z_2$ gauge theory and a two-form $\Z_2$ gauge theory. It is instructive to see how this fails. The flux-attached Gauss constraint depends on the twist fields, as shown in eq.\ \eqref{2d flux attached twisted gauge constraint}, and it must be promoted into
\bel{\label{2d alternative gauge constraint 0}
  \1 = \~G_f^\vee \sideset{}' \prod_{v \subset f} Z_v^\vee = - (\delta\eta)_f \prod_{\ell \subset f} \~X_\ell^\vee \sideset{}' \prod_{v \subset f} \left(W_v^\vee Z_v^\vee\right).
}
The one-form gauge constraint will be discussed below, but for now let
\bel{
  \~W_v^\vee \equiv W_v^\vee Z^\vee_v
}
be the tentative one-form-gauge-invariant version of the Wilson loop. (This object dualized to $Z_v$ in the KW version of this story, eq.~\eqref{2d KW alternative 2}.) In this notation the zero-form Gauss law \eqref{2d alternative gauge constraint 0} is
\bel{\label{2d alternative gauge constraint 1}
  \widehat G^\vee_f \equiv \prod_{\ell \subset f} \~X^\vee_\ell \sideset{}' \prod_{v \subset f} \~W_v^\vee = - (\delta \eta)_f.
}
Recall that the r.h.s.~is fixed by requiring consistency of the standard JW duality \eqref{2d JW twisted}.

Notice that now that the one-form gauge fields are present, the notion of flux attachment in the ordinary gauge theory is obscured: the original flux-attaching Gauss operator \eqref{2d flux attached gauge constraint} is no longer gauge-invariant. The meaningful gauge constraint that supplants this old Gauss law is eq.~\eqref{2d alternative gauge constraint 1}. As already mentioned, it is impossible to get rid of the background magnetic fields $(\delta\eta)_f$ on a generic lattice. However, this is not the anomaly that prevents the twists $\eta^\vee_v$ from being quantized.

The obstacle lies in the fermionic nature of the one-form symmetry generated by line operators $\~T_c^\vee$ on closed loops. In order to gauge this symmetry, a set of local Gauss operators $\widehat G^\vee_\ell$ must be associated to the set of links on $\Mbb^\vee$. The one-form Gauss operator in \eqref{2d one-form gauge constraint} needs to be modified in order to get a meaningful gauge constraint. It cannot be used outright because $X_\ell^\vee$ is not gauge-invariant under the zero-form gauge symmetry (i.e.~$X_\ell^\vee$ does not commute with $\widehat G_f^\vee$), and the natural generalization, $\~X_\ell^\vee X^\vee_{\ell_1} X^\vee_{\ell_2}$, cannot be used because such operators on different links do not necessarily commute with each other.\footnote{These operators also do not commute with the zero-form Gauss operator $\widehat G^\vee_f$ in \eqref{2d alternative gauge constraint 1}, but since $\widehat{G}^\vee_f$ must be realized as a product of one-form Gauss operators along the dual of $\del f$, this failure of commutativity simply follows from the failure of different $\~X_\ell^\vee$ to commute.} Further modifications must be done by adding various factors of $Z_v^\vee$ to $\~X_\ell^\vee X^\vee_{\ell_1} X^\vee_{\ell_2}$, getting tentative one-form Gauss operators
\bel{\label{temp}
  \widehat G^\vee_\ell = \~X_\ell^\vee \prod_{v \subset \ell} X_{v}^\vee (Z_v^\vee)^{\alpha_v(\ell)}, \quad \alpha_v(\ell) \in \Z_2.
}
If $\ell$ is in the boundary $\del \Mbb^\vee$, there are two options. It can have no dual faces adjacent to it, in which case it can be excluded from this analysis. Alternatively, $\ell$ can have one dual face adjacent to it, in which case the corresponding one-form Gauss operator is assumed to be $\widehat G^\vee_\ell \propto X_\ell^\vee X_{\ell_1}^\vee (Z_{\ell_1}^\vee)^{\alpha(\ell)}$, where $\ell_1$ denotes the single dual face containing $\ell$. For simplicity, the rest of this argument assumes that $\Mbb^\vee$ has no boundary.

If the dual lattice $\Mbb^\vee$ is triangular, it is possible to choose the $\widehat G^\vee_\ell$'s such that they all commute. However, it is never possible to make them individually commute with the zero-form Gauss operators. Even a more modest goal, to make the one-form operators consistent with the zero-form Gauss law via vertex relations, is also impossible.

The claim that operators \eqref{temp} cannot commute with $\widehat G^\vee_f$ is quickly established by inspection. A straightforward way to prove the rest of the claims from the above paragraph is as follows. Consider any dual lattice $\Mbb^\vee$ with some fixed definition of $\~X_\ell^\vee$'s. The two requirements of interest are (i) that any two operators $\widehat G^\vee_\ell, \widehat G^\vee_{\ell'}$ commute, and (ii) that the $\widehat G^\vee_\ell$'s be consistent with the zero-form Gauss law up to a sign by satisfying the vertex relation $\prod_{\ell \subset f} \widehat G_\ell^\vee \propto \widehat G_f^\vee$.\footnote{%
For every dual link $\ell$, the operators that may fail to commute with $\~X^\vee_\ell$ are $\~X^\vee_{\ell'}$ for $\ell' \subset \ell_{1/2}$; the ansatz \eqref{temp} guarantees that when $\ell$ and $\ell'$ do not share a dual face, the corresponding Gauss operators $\widehat G^\vee_\ell$ and $\widehat G^\vee_{\ell'}$ commute.%
}
Can these requirements be satisfied? Consider a dual vertex $f$ with $z(f)$ dual links emanating from it. By req.~(i), for any two links $\ell, \ell' \subset f$ that share a dual face, $\widehat G^\vee_\ell$ and $\widehat G^\vee_{\ell'}$ must commute; this gives $z(f)$ constraints on the $\alpha_v(\ell)$'s. This is done independently at each $f$, so there are $\sum_f z(f) = 2N\_L$ constraints of this type, with $N\_L$ being the number of dual links in $\Mbb^\vee$. This is precisely enough to fix all the $\alpha_v(\ell)$'s. There are more constraints, however: it is simple to check that no req.~(ii) can be satisfied with such $\alpha_v(\ell)$'s. Moreover, if the lattice $\Mbb^\vee$ is not triangular, req.~(i) is not exhausted yet either: one still needs to enforce the commutation of the $\widehat G^\vee_\ell$'s on dual links that belong to the same dual face but that do not emanate from the same dual vertex. This proves both statements from the preceding paragraph.

Note that this one-form anomaly is a property of the algebra, not of any particular Hamiltonian. Any zero-form gauge theory with flux attachment, or more precisely any theory with a framing-dependent one-form symmetry, will run into the same issue: not all $\widehat G^\vee_\ell$'s can be simultaneously projected to the singlet sector without violating the zero-form gauge symmetry. The next best alternative is to partially gauge the one-form symmetry, i.e.~to project to the singlet sector of only those operators $\widehat G^\vee_\ell$ that are zero-form gauge-invariant and that commute with each other. These must live on links $\ell$ whose adjacent dual faces $\ell_{1/2}$ do not contain anchors on $\ell$; this means that an $O(1)$ fraction of links may not able to support the desired $\widehat G^\vee_\ell$'s. Such theories would thus need to be dual to fermions coupled to an entire additional field theory. This analysis lies outside of the scope of this paper, but see \cite{Bhardwaj:2016clt} for further work in this direction.
\bigskip

To recap, JW in $d = 2$ involves a more limited web of dualities due to the associated anomalous one-form symmetries. Two dualities of interest are
\boxedAlgns{
  \trm{fermions}^\eta/\Z_2 &= \Z_2\ \textrm{flux-attached\ gauge\ theory}^\vee/(\Z_2\ \textrm{one-form}),\\
  \trm{topologically\ gauged\ fermions}^\eta &= \Z_2\ \textrm{flux-attached\ gauge\ theory}^\vee.
}
The superscript $\eta$ is a reminder here that fermions may need to be coupled to a nontrivial background $\Z_2$ connection $\eta_\ell$ to make the duality consistent, even if the fermionic theory is well-defined on its own. Once $\eta_\ell$ is promoted to a dynamical gauge field, the duality is well-defined if the gauge field is topological with fixed curvature $(\delta \eta)_f$.

Unlike in $d = 1$, here there is no natural theory to use as a reference point when defining the boundary conditions of fermions. A different choice of the background gauge field $\eta_\ell$ on the fermion side can change what one means by boundary conditions. Nevertheless, the notion of summing over boundary conditions remains perfectly well-defined: this is precisely the procedure of promoting the background fields $\eta_\ell$ into quantum degrees of freedom. In any $d > 1$, this sum over boundary conditions will always be equivalent to coupling to a $\Z_2$ gauge field in the usual way, as done in \eqref{2d JW alternative}. It is only in $d = 1$ that the sum over fermion boundary conditions may require the unusual Gauss law \eqref{1d flux attached gauge constraint} and the correspondingly unusual map \eqref{1d JW original 2}.

\subsection{Comments on orientations, dimers, and spin structures} \label{subsec 2d spin structures}

It is well known that boundary conditions along noncontractible cycles for fermions on a space $\bb X$ are closely associated to spin structures on $\bb X$.\footnote{Note that $\Mbb$ is always used to denote the spatial manifold or lattice, while $\bb X$ may also be a spacetime.} This connection has several manifestations on both continuous and discrete spaces. The classic physical approach to spin structures comes from relativistic quantum field theory on an orientable $D$-dimensional manifold $\bb X$, where representations of the Lorentz group can be spinorial. A field in such a representation is a section of a spinor bundle. This spinor bundle is obtained by uplifting from the tangent bundle, with a $\Z_2$ connection specifying the additional data needed for this uplift. (In Euclidean signature, this is a lift from $\SO(D)$ to $\trm{Spin}(D)$ connections.\footnote{Note that $\pi_1(\SO(D)) = \Z_2$ for $D > 2$, while $\pi_1(\SO(2)) = \Z$. In any $D$, $\Spin(D)$ is such that $\SO(D) = \Spin(D)/\Z_2$, but only for $D = 2$ this means $\Spin(2) = \SO(2)$. The subtlety of spin connections in $D = 2$ vs.~$D > 2$ will be discussed at the end of this subsection, and also in subsection \ref{subsec paraspin structures}}.) This extra piece of data is the spin structure, and on a smooth manifold its only detectable signature is a $\Z_2$-valued holonomy along noncontractible cycles. These holonomies can be interpreted as periodic or antiperiodic boundary conditions for spinors governed by a Dirac Lagrangian. According to the spin-statistics connection, these spinors must be fermions, and this is how fermions get associated to spin structures.

This familiar story relies on $\bb X$ being a (pseudo-)Riemannian manifold. Spin structures, however, are $\Z_2$ connections and can be naturally defined on lattices. One way to do so is to start from the discrete version of the second Stiefel-Whitney class, which is a certain $\Z_2$ two-cocycle $w_2$ on the lattice $\bb X$ for which very explicit formul\ae\ are given in \cite{Goldstein:1976, Halperin:1972}.\footnote{A few warnings on conventions: the Stiefel-Whitney classes used in this paper are all $\Z_2$ ones. The \emph{integral} Stiefel-Whitney classes are used to define spin$^c$ structures but will not be explored here. Further, the given references define Stiefel-Whitney classes as cycles, not cocycles, so their definitions need to be appropriately (Poincar\'e-)dualized. Finally, the second Stiefel-Whitney class is typically defined as a $\Z_2$ two-cocycle, meaning that it is a function assigning an element of $\Z_2 = \{0,1\}$ to each face. However, it is more convenient to work with its exponential taking values in $\{\pm 1\}$, and this is what $w_2$ will mean in this paper.} The spin structure $\xi$ is then any one-cochain that satisfies $w_2 = \delta \xi$. (If $w_2$ is not exact, then $\bb X$ does not admit spin structures; this does not happen to orientable lattices in $D \leq 3$.) This $\xi$ can be regarded as a classical $\Z_2$ gauge field, and once $w_2$ is fixed, the only ``gauge-invariant'' data contained in $\xi$ are precisely the values of its holonomies along noncontractible cycles.

If $\bb X$ is a lattice in two dimensions, i.e.~if it can be smoothly embedded into a 2D surface, then spin structures can also be defined via $\Z_2$ quadratic forms on one-cycles following \cite{Atiyah:1971, Johnson:1980, Cimasoni:2008}. The idea here is to define $\Z_2$-valued functions $q_\xi(c)$ that act on one-cycles $c \in \bb X$ and satisfy
\bel{\label{def q xi}
  q_\xi(c + c') = q_\xi(c) + q_\xi(c') + c \cdot c',
}
where $c \cdot c'$ is the bilinear intersection form that measures the number of intersections (mod 2) of the two chains $c, c'$. The space of such forms $q_\xi(c)$ is $2^{2g}$-dimensional, where $g$ is the genus of $\bb X$. The distinct basis labels $\xi$ correspond to nonequivalent spin structures; for each $\xi$, $q_\xi$ assigns a $\Z_2$ variable to each noncontractible cycle.

Having learned about discrete spin structures, one may ask whether they may be connected to fermionic quantum theories on lattices. Remarkably, a connection between fermionic theories and spin structures persists even in a discrete setup. There are two different avatars of this connection in the literature, and they correspond to the two definitions of discrete spin structures given above.

The relation between fermions and spin structures as quadratic forms on 2D surfaces has an interesting background. The oldest chapter of this story is a set of dualities established within statistical mechanics by Kasteleyn and Fisher \cite{Fisher:1966, Kasteleyn:1963}. In modern words, these dualities relate Ising spins on a 2D lattice $\~{\bb X}$ to dimer models%
\footnote{\label{foot dimers}A dimer model on a lattice $\bb X$ has as its degrees of freedom different \emph{perfect matchings} of adjacent vertices on $\bb X$. In other words, each configuration is a way of assigning $0$ or $1$ to the links in $\bb X$ such that each vertex has exactly one $1$-link (dimer) emanating from it. Each link $\ell\in \bb X$ is assigned a weight $w(\ell)$, and the weight of each dimer configuration (set of $1$-links) $\mathfrak D$ is simply $w(\mathfrak D) \equiv \prod_{\ell \in \mathfrak D} w(\ell)$, giving the partition function $\mathcal Z = \sum_{\mathfrak D} w(\mathfrak D)$.} %
on an associated lattice $\bb X$, which are in turn related to spinless fermions hopping on $\bb X$ in the presence of an unusual topological gauge field $w_{\mathfrak K}$, to be described below. According to the prescription of \cite{Fisher:1966}, if $\~{\bb X}$ has a time direction (e.g.~it is a discretization of the two-torus or a plane), $\bb X$ does not have a corresponding time direction, so it may appear that this duality has no interpretation in the Hamiltonian framework, despite the natural guess that it is just a Euclidean version of \eqref{1d JW original 2}. The unusual nature of gauge fields that couple to fermions on $\bb X$ appears to restore a Hamiltonian interpretation to this theory in this case.

More recent is Kuperberg's connection between dimer configurations on the 2D lattice $\bb X$ and spin structures on surfaces in which $\bb X$ may be embedded \cite{Kuperberg:1998}. Combining this connection with the dimer-fermion duality and the definition of spin structures as quadratic forms \eqref{def q xi}, Cimasoni and Reshetikhin showed that a dimer configuration $\mathfrak D$ and an appropriate (``Kasteleyn'') choice of link orientations $\mathfrak K$ canonically induce a $\Z_2$ quadratic form $q_{\mathfrak D, \mathfrak K}$ on $\bb X$ \cite{Cimasoni:2008}. This gives a precise way to identify the topological gauge fields/Kasteleyn orientations with spin structures, and the statistical mechanics of dimers is then dual to that of fermions coupled to dynamical spin structures. Combining this with the Ising-dimer duality, this gives the map between Ising spins and ``gauged'' fermions:
\bel{\label{dimer Z}
  \mathcal Z\_{Ising} \propto \sum_{\mathfrak K} \eps_{\mathfrak{D,\, K}} \ \trm{Arf}(q_{\mathfrak{D,\, K}}) \int \d \chi \, \exp\bigg\{\sum_{\avg{u,v}} w_{\mathfrak K}(u,v) \chi_u \chi_v\bigg\}.
}
Here $\mathfrak K$ labels different (possibly gauge-equivalent) spin structures/Kas\-te\-leyn orientations/``gauge field'' configurations, $\eps_{\mathfrak{D,\, K}}$ is a sign whose explicit dependence on $\mathfrak{D, K}$ is given in \cite{Cimasoni:2008}, and the Arf invariant is a standard function on $\Z_2$ quadratic forms in 2D \cite{Arf:1941}. The summand is independent of $\mathfrak D$ and is invariant under ``gauge transformations'' that change the orientation $\mathfrak K$ of all links emanating from one site.\footnote{The need for a fiducial $\mathfrak{D}$ becomes more natural in view of the fact that lattice spinors are naturally defined via a perfect matching of spinless fermions \cite{Susskind:1976jm}. A similar observation was made in \cite{Dijkgraaf:2007yr}.} The integral is the standard Berezin path integral. Its evaluation yields the Pfaffian of the antisymmetric matrix $w_{\mathfrak K}$, whose nonzero entries are the ``gauge fields''
\bel{
  w_{\mathfrak K}(\ell_1, \ell_2) = - w_{\mathfrak K}(\ell_2, \ell_1) \equiv w(\ell),
}
where $w(\ell)$ are the link weights from the dimer model (see footnote \ref{foot dimers}).  This is an unusual gauge field because it depends on the orientation; it is topological because its action, $\eps_{\mathfrak{D,\, K}} \, \trm{Arf}(q_{\mathfrak{D,\, K}})$, does not depend on the field strength but only on the holonomies along noncontractible cycles. Note that this property is built into the definition of $\mathfrak K$: an orientation is Kasteleyn if its ``curvature'' (i.e.~the product of $w_{\mathfrak K}(u,v)$ over a one-cycle bounding a face) is equal to $-1$.\footnote{With some extra assumptions on the lattices involved, Cimasoni has further shown how to extend this story to Dirac fermions on 2D lattices \cite{Cimasoni:2009}.}

Another avatar of the nonrelativistic link between spin structures and fermions is well represented in the recent work by Kapustin and collaborators \cite{Bhardwaj:2016clt, Kapustin:2017jrc, Gaiotto:2015zta}. Compared to the ideas mentioned in the previous passage, these authors have studied a more straightforward set of KW and JW dualities in topological field theories, both via Hamiltonian methods and discrete path integrals (``state sums''). Promoting twist fields to quantum variables in JW dualities was called ``gauging fermion parity'' (for zero-form symmetries) and ``anyon condensation'' (for one-form symmetries). In $D = 2$, a picture similar to Kasteleyn's was found, with the Arf invariant appearing in the action of spin structures when they were dynamical \cite{Kapustin:2017jrc}. In $D \geq 3$, the information on the spin structure was inferred from the anomaly of the one-form symmetry, i.e.~by studying which additional degrees of freedom need to be introduced in order to get a consistent one-form-gauged theory \cite{Bhardwaj:2016clt}. By carrying out the analogue of the state sum approach in the Hamiltonian framework, it was shown that the notion of spin structures defined via discrete Stiefel-Whitney classes of the spatial manifold coincides with the information contained in the structure of one-form anomalies. The twist \eqref{2d JW alternative} of the standard JW duality  \eqref{2d JW twisted} in the present paper is the dual of this story: here the notion of summing over spin structures is uncovered via gauging the nonanomalous zero-form symmetry of fermions, not the one-form symmetry of the dual flux-attached gauge theory.

Several lessons about the connection between spin structures and $\Z_2$ gauge fields can be drawn from the past two sections and from this blitz review of previous results:
\begin{enumerate}
  \item In all dimensions, a spin structure can be realized as a topological gauge field whose gauge group is the $\Z_2$ fermion parity. It can be defined in any theory in which the operator algebra is generated by fermion bilinears, i.e.~in which fermion parity is a kinematic symmetry. (It is not necessary to assume the theory is topological, as in \cite{Bhardwaj:2016clt}.) If the spin structure is a classical (background) field, the theory can be called a \emph{spin theory}. If it is a quantum (dynamical) field, the theory is a \emph{non-spin theory}. This nomenclature is consistent with the definition of spin topological field theories.
  \item In a spin theory, the dependence on the spin structure does not enter at the level of the operator algebra. In fact, the spin structure is not distinguishable from the couplings in the Hamiltonian. Only changes in the spin structure can be defined unambiguously --- this is why the space of spin structures is affine. One way to define what one means by a theory with a specific spin structure is to pick a reference Hamiltonian built out of fermion bilinears.  For example, if the reference Hamiltonian is $\sum_\ell \eta_\ell S_\ell$, then changing the $\eta_\ell$'s while keeping $(\delta\eta)_f$ fixed will give a theory with distinct boundary conditions --- i.e.~a different spin structure --- if the $\eta_\ell$'s are changed in a topologically nontrivial way (otherwise the changes can be absorbed by a phase shift of the matter operators). Another way to define a reference Hamiltonian is via duality: as shown in this paper, different spin structures can be identified with different singlet/singlet dualities. Thus in $d = 1$ one can pick, say, the Ising Hamiltonian \eqref{def Ising} in the singlet sector and dualize it (with $\eta_\ell = 1$) to get a reference Hamiltonian for the fermionic system.
  \item The space of spin structures is spanned by holonomies $\prod_{\ell \subset c} \eta_\ell$ over noncontractible cycles, i.e.~by gauge-inequivalent fields $\eta$ that all have the same field strength $\delta \eta$. This is very reminiscent of the definition of spin structures $\xi$ as inequivalent solutions to $\delta \xi = w_2$. These two approaches can in fact coincide if the conventions from subsection \ref{subsec 2d JW} are adjusted in the following way. Recall that the vertex relation \eqref{2d vertex relation} can be consistent with $(\delta \eta)_f = 1$ as long as the dual $\Z_2$ theory with flux-attachment is endowed with a nontrivial background charge density, with some Gauss constraints in \eqref{2d flux attached gauge constraint} being $\~G^\vee_f = -\1$. If this background charge is included in the definition of the gauge theory, the equation $(\delta \eta)_f = 1$ will then be the same (up to an irrelevant overall sign) as the equation obeyed by $\xi$, as the second Stiefel-Whitney class on any orientable 2D lattice is the constant two-cocycle, $(w_2)_f = -1$. In higher dimensions, more interesting choices of $w_2$ will be possible, but the discussion is the same as here \cite{Chen:2018nog}.
  \item In the Hamiltonian framework, $d = 1$ is a special dimension. Trivially, there is no definition of spin structures via Stiefel-Whitney classes here. More interestingly, though, only in this case can a $\Z_2$ gauge field be traded (via gauge-fixing) for a single $\Z_2$ degree of freedom, so a theory whose algebra has individual fermionic operators (not just fermion bilinears) can be interpreted as a non-spin theory (this is how the original JW duality is understood in \eqref{1d JW original 2}). Further, spin structures in $d = 1$ can be promoted to dynamical $\Z_2$ gauge fields but their Gauss constraint may be of the unusual form (cf.~\eqref{1d flux attached gauge constraint}) that includes nonlocal holonomies.
  \item Correspondingly, in statistical mechanics/state sum constructions, $D = 2$ is a special dimension. There is again a certain leeway in defining spin structures. They can be defined as usual topological $\Z_2$ gauge fields like in eq.~\eqref{1d JW alternative}, or as unusual topological gauge fields (Kasteleyn orientations). In this ``unusual'' case they enter the partition function weighted by the Arf invariant, cf.~\eqref{dimer Z}. It would be interesting to understand whether the unusual Gauss constraints in $d = 1$ and unusual gauge fields in $D = 2$ are connected by a conceptual link that the Arf term in the action hints at.
  \item In higher dimensions, spin structures are ordinary topological $\Z_2$ gauge fields, background or dynamical, that couple to fermions. In state sum constructions of non-spin theories, their actions are ordinary weak coupling Kogut-Susskind actions \cite{Kogut:1974ag}, perhaps with background fluxes turned on as required by the vertex relations analogous to \eqref{2d vertex relation}, but without any dependence on nonlocal operators (holonomies) that the Arf invariant had introduced in $d = 1$.
\end{enumerate}

\section{$\Z_K$ dualities} \label{sec Zk}

All dualities in the previous section involved theories with $\Z_2$ target spaces. This class of theories is particularly natural because it includes pure fermion theories in any dimension. However, all duality webs discussed so far admit a generalization to arbitrary $\Z_K$ theories. JW-type dualities now become maps between bosonic and \emph{parafermionic} theories, which are less familiar --- in part because they exhibit less interesting critical phenomena above $d = 1$, largely as a result of constraints due to Lorentz symmetry and the spin-statistics relation. On the other hand, KW maps remain of great interest in these cases, especially in the $K \rar \infty$ limit which reveals particle-vortex and related dualities.

Consider a bosonic matter theory with $K$ degrees of freedom per site --- a \emph{clock model}. Let
\bel{
  \omega \equiv \e^{2\pi \i/K}
}
and note that $\omega^K = 1$. The generalizations of Pauli matrices $Z$ and $X$ to $K > 2$ are
\bel{\label{def Phi Pi}
  \Phi = \left[
             \begin{array}{cccc}
               1 &  &  &  \\
                & \omega &  &  \\
                &  & \ddots &  \\
                &  &  & \omega^{K - 1} \\
             \end{array}
           \right],
  \quad
  \Pi = \left[
            \begin{array}{cccc}
                &  &  & 1 \\
               1&  &  &  \\
                & 1 &  &  \\
                &  & \ddots &  \\
            \end{array}
          \right].
}
They obey $\Phi^n \Pi^{m} = \omega^{n m} \Pi^{m} \Phi^n$, $\Phi^{-1} = \Phi\+ = \Phi^{K - 1}$, $\Pi^{-1} = \Pi\+ = \Pi^{K - 1}$. A typical Hamiltonian is
\bel{\label{def H K}
  H_K = \frac1{2M} \sum_v \left(\Pi_v + \Pi^{-1}_v\right) + \frac12 \sum_{\avg{u, v}} \left(\Phi^{-1}_u \Phi_v + \Phi^{-1}_v \Phi_u \right) + \sum_v V\left(\frac12(\Phi_v + \Phi_v^{-1})\right).
}
It is often useful to let $\Phi_v \equiv \e^{\i \phi_v}$ and to assume that, at $K \gg 1$, states with slowly varying $\phi_v$ eigenvalues form the low energy effective theory. Then the above model can be approximated by the familiar compact scalar Lagrangian
\bel{
  H_{K \gg 1} \approx \frac1{2M} \sum_v \dder{}{\phi_v} - \frac12 \sum_{\avg{u, v}} (\phi_u - \phi_v)^2 + \sum_v V(\cos\phi_v).
}
Deep in the IR, the fluctuations of the scalar field $\phi$ may be small, in which case the model is well described by the most familiar scalar field theory, the noncompact (real-valued) scalar field.

Gauge theories with a $\Z_K$ group can be defined using operators $\Phi_\ell$, $\Pi_\ell$ from \eqref{def Phi Pi} on links, but this time orientations are important even without flux attachment. As before, $\ell_1, \ell_2$ denote the vertices in $\ell$ such that the link is oriented from $\ell_1$ to $\ell_2$. A useful notation is
\bel{
  \Phi_{\ell_1, \ell_2} = \Phi^{-1}_{\ell_2, \ell_2} \equiv \Phi_\ell,
}
and the same for $\Pi_\ell$. In addition, $\Phi_{u, v} \equiv \1$ and $\Pi_{u, v} \equiv \1$ if $u, v$ do not belong to the same link. The Gauss operators in the pure $\Z_K$ gauge theory are
\bel{\label{def G Zk}
  G_v = \prod_u \Pi_{v, u}.
}
Generators of the gauge-invariant algebra are $\Pi_\ell$ and the Wilson loops
\bel{
  W_f = \prod_{i = 1}^{|f|} \Phi_{v_i, v_{i + 1}},
}
where the vertices $\{v_i\}$ in $f$ are ordered counter-clockwise, and $|f|$ is the number of vertices in $f$. Of course, if the lattice has noncontractible cycles $c$, Wilson loops $W_c$ must be added to the set of generators.

Just like in the scalar case, it is often useful to let $\Phi_\ell \equiv \e^{\i A_\ell}$ and to consider states where $A_\ell$ varies slowly across the lattice. At $K \gg 1$ this gives back the compact U(1) gauge theory. (More precisely, to get U(1) from $\Z_K$, $K$ must be taken to infinity first, before the large lattice or weak coupling limits \cite{Radicevic:2015sza}.)

Finally, parafermion analogues $\zeta, \xi$ of Majorana operators $\chi, \chi'$ are defined such that
\bel{\label{def paras}
  \zeta_v^K = \xi_v^K = \1,\quad \zeta_v\+ = \zeta^{-1}_v, \quad \xi_v\+ = \xi^{ - 1}_v
}
and
\gathl{\label{para comm rels}
  \zeta_u \zeta_v = \omega \zeta_v \zeta_u, \quad \xi_u \xi_v = \omega \xi_v \xi_u, \quad \trm{for}\quad u < v,\\
  \zeta_u \xi_v = \omega \xi_v \zeta_u \quad \trm{for} \quad u \leq v, \qquad \qquad \zeta_u \xi_v = \omega^{-1} \xi_v \zeta_u, \quad \trm{for} \quad u > v.
}
Here an absolute ordering of the vertices has been chosen. Note that a simpler relation, $\zeta_u \zeta_v = \omega \zeta_v \zeta_u$ for all $(u, v)$, cannot simultaneously hold for both $(u, v)$ and for $(v, u)$, because $\omega \neq \omega^{-1}$ for $K > 2$. In $d = 1$, it is possible to let $\xi_v \equiv \zeta_{v + \frac12}$, unifying the relations \eqref{para comm rels} into just $\zeta_u \zeta_v = \omega \zeta_v \zeta_u$ for $u < v$. For more details on parafermions in $d = 1$, see \cite{Fendley:2013snq}.

\subsection{Kramers-Wannier and its twisting} \label{subsec KW Zk}

The standard KW dualities \eqref{1d KW original} and \eqref{2d KW original} remain intact as $\Z_2$ is generalized to $\Z_K$. However, in order to make contact with familiar continuum notions, it is useful to consider the ``conjugate'' dualities obtained by swapping position and momentum operators in the $\Z_2$ expressions. In $d = 1$, the map of interest is
\algns{\label{1d KW Zk}
  \Pi_v &= (\Phi^\vee_{v - \frac12})\+ \Phi^\vee_{v + \frac12},\\
  \Phi\+_{v - 1} \Phi_v &= \Pi_{v - \frac12}^\vee.
}
Consistency requires
\bel{
  Q \equiv \prod_v \Pi_v = \1, \quad Q^\vee \equiv \prod_v \Pi_v^\vee = \1,
}
where $Q$ and $Q^\vee$ generate a global $\Z_K$ symmetry. The Hamiltonian \eqref{def H K} is invariant under this symmetry only if $V = 0$; at $K \rar \infty$ this becomes the U(1) shift symmetry $\phi_v \mapsto \phi_v + 2\pi/K$ familiar from the case of the free scalar. The duality can be written as
\algns{
  -\i\der{}{\phi_v} &= \phi_{v + \frac12}^\vee - \phi_{v - \frac12}^\vee,\\
  \phi_v - \phi_{v - 1} &= -\i \der{}{\phi^\vee_v}.
}
In the more common continuum path integral notation in $D = 2$, in a specific convention for directions of derivatives, this duality becomes
\bel{\label{1d KW Zk cont}
  \del^\mu \phi = \epsilon^{\mu\nu} \del_\nu \phi^\vee.
}
Note that \eqref{1d KW Zk cont} holds only at long distances and with $K \rar \infty$ taken first, while the duality \eqref{1d KW Zk} holds exactly on the lattice, for any $K$. Both are self-dualities, just like \eqref{1d KW original}.

The singlet condition $Q = \1$ means that the zero-momentum mode $\phi_0 \equiv \frac1N\sum_v \phi_v$ is not dynamical. It is the only mode that changes under the shift symmetry, and so $Q = \1$ forces all physical states to be singlet eigenstates of $\phi_0$, i.e.~tensor products of arbitrary nonzero mode states with the zero mode state $\frac1{\sqrt K} \sum_{n = 1}^K \qvec{\frac{2\pi n}K}$.

This map can be twisted by $\eta_{v - \frac12}, \eta_v^\vee \in \{1, \omega, \ldots, \omega^{K - 1}\}$, getting
\algns{\label{1d KW Zk twisted}
  \Pi_v &= \eta^\vee_v (\Phi^\vee_{v - \frac12})\+ \Phi^\vee_{v + \frac12},\\
  \eta_{v - \frac12} \Phi\+_{v - 1} \Phi_v &= \Pi_{v - \frac12}^\vee,
}
with consistency equations $Q = \prod_v \eta^\vee_v$ and $Q^\vee = \prod_v \eta_{v + \frac12}$. Promoting, say, $\eta_v^\vee$ to a dynamical $\Z_K$ gauge field, this becomes a duality between a gauged scalar theory and a full (non-singlet) scalar theory,
\algns{\label{1d KW Zk alternative}
  \Pi_v &= (\Phi^\vee_{v - \frac12})\+ \Phi_v^\vee \Phi^\vee_{v + \frac12},\\
  \Phi_v\+ &= \Pi_v^\vee,
}
with the usual gauge constraint $G^\vee_{v + \frac12} = (\Pi^\vee_v)\+ \Pi_{v + \frac12}^\vee \Pi_{v + 1}^\vee$. (Note the appearance of $\Phi\+_v$ in the second line.) In the continuum notation analogous to \eqref{1d KW Zk cont}, still keeping lattice spacings equal to unity, this is
\algns{\label{1d KW Zk alt cont}
  \del_\mu \phi &= \epsilon^{\mu\nu}(A_\nu^\vee + \del_\nu \phi^\vee),\\
  \phi &= \epsilon^{\mu\nu} F_{\mu\nu}^\vee.
}

Finally, there is an interesting extra twist to the twisting story here: the twist fields can come with various charges. Consider twisting \eqref{1d KW Zk} into
\algns{\label{1d KW Zk twisted 2}
  \Pi_v &= \big(\Phi^\vee_{v - \frac12}\big)\+ \Phi^\vee_{v + \frac12},\\
  \eta_{v - \frac12}^q \Phi\+_{v - 1} \Phi_v &= \Pi_{v - \frac12}^\vee.
}
The singlet constraints are now $Q = \1$ and $Q^\vee = \prod_v \eta_{v - \frac12}^q$. Upon promoting the $\eta$'s to gauge fields, the second constraint becomes
\bel{\label{1d ZK orbi constraint}
  W^q = Q^\vee,
}
and the gauge constraint is
\bel{
  G_v \equiv  \Pi\+_{v - \frac12}\, \Pi^q_v\, \Pi_{v + \frac12} = \1.
}
A theory where matter has charge $q$ will be called a \emph{$q$-gauged theory}. Note that the Gauss law implies the constraint
\bel{
  Q^q = \1,
}
but the first line of the duality \eqref{1d KW Zk twisted 2} still implies
\bel{
  Q = \1.
}
If $K = qr$ for some integer $r$, then the gauged clock model has a gauge group ``broken'' to $\Z_r$ \cite{Fradkin:1978dv}. More precisely, the matter only couples to a $\Z_r$ subgroup of the gauge field, and the remaining $\Z_q$ symmetry of the clock model is simply projected to the singlet sector to ensure the consistency of the duality. The dual theory is also in the singlet sector of the $\Z_q$ subgroup; here the $q$'th power of individual $\Phi^\vee_{v + \frac12}$ operators now becomes physical, giving the duality
\algns{\label{1d KW Zk orbifold}
  \Pi_{v + \frac12} &= \big(\Phi_{v + \frac12}^\vee\big)^q,\\
  \Pi_v &= \big(\Phi^\vee_{v - \frac12}\big)\+ \Phi^\vee_{v + \frac12},\\
  \Phi\+_{v - 1} \Phi^q_{v - \frac12} \Phi_v &= \Pi_{v - \frac12}^\vee.
}
The r.h.s.~can be interpreted as $q$ copies of a $\Z_r$ clock model with effective position operators $\big(\Phi^\vee_{v + \frac12}\big)^q$, momentum operators $\Pi^\vee_{v + \frac12}$, and \emph{orbifold (twist) operators} $\Phi^\vee_{v + \frac12}$ --- essentially $q$'th roots of the effective  position operators, whose action is to shift from one copy of the theory to the other. The charge $Q^\vee = \prod_v \Pi_{v + \frac12}^\vee$ combines both the $\Z_r$ and $\Z_q$ charges of this system; the $\Z_q$ part (the ``replica symmetry'') is generated by $(Q^\vee)^r$, and by \eqref{1d ZK orbi constraint} this symmetry must be in its singlet sector: the r.h.s.~is really a $\Z_q$ orbifold of a $\Z_r$ clock model. The $\Z_r$ charge (the shift symmetry of the effective $\Z_r$ clock model) is fully dynamical.

To summarize, $d = 1$ lattice dualities exhibited so far are
\boxedAlgns{
  \Z_K \trm{\ clock\ model}/\Z_K &= \Z_K \trm{\ clock\ model}^\vee/\Z_K,\\
  \Z_K \trm{\ clock\ model} &= \trm{gauged}\ \Z_K \trm{\ clock\ model}^\vee,\\
  q\textrm{-gauged\ }\Z_K\trm{\ clock\ model}/\Z_q &= \Z_q\textrm{-orbifolded\ } Z_{K/q} \trm{\ clock\ model}^\vee.
}
In the continuum, as $K \rar \infty$, the dualities discussed above were
\boxedAlgns{
  \trm{compact\ scalar/U(1)} &= \trm{compact\ scalar}^\vee/\trm U(1),\\
  \trm{compact\ scalar} &= \trm{gauged\ compact\ scalar}^\vee.
}
Note that more dualities can be unearthed in straightforward ways: for instance, the continuum models also admit dualities of orbifolded compact scalars.

\bigskip

In $d = 2$, the situation is again a straightforward generalization of the $\Z_2$ analysis. The standard duality is between a $\Z_K$ scalar theory and a pure $\Z_K$ gauge theory with Gauss operators \eqref{def G Zk},
\algns{\label{2d KW Zk original}
  \Phi_{\ell_1}\+ \Phi_{\ell_2} &= \Pi_\ell^\vee, \\
  \Pi_v &= W^\vee_v.
}
Note which of the two $\Phi$'s enters the duality conjugated. This choice follows from the rule that link orientations on $\Mbb^\vee$ are induced by a $\pi/2$ \emph{clockwise} rotation of the orientation of links on $\Mbb$. When $K \rar \infty$, these duality relations at long distances can be written as
\algns{\label{2d KW cont}
  \del^\mu \phi = \epsilon^{\mu\nu\lambda} F_{\nu\lambda}^\vee.
}
This is the familiar particle-vortex duality.\footnote{More precisely, as mentioned in the introduction, this is the compact/``weak'' part of the particle-vortex duality \cite{Senthil:2018}. There has recently been a flurry of activity investigating the noncompact particle/vortex dualities, in which the scalar field on at least one side of the duality is not compact. These dualities are typically not exact and will not be addressed here.} Note that it holds only for theories whose Hamiltonians do not contain $\phi$ operators, as those would clash with the requirement that the scalar theory be in the singlet sector of the shift symmetry. The dual gauge theory is in the singlet sector of a one-form $\Z_K$ symmetry generated by analogues of $T^\vee_c$ in \eqref{def T}. In any $d$, the above duality generalizes to a map between a zero-form $\phi$ and a $(d - 1)$-form $B_{(d - 1)}^\vee$ given by
\bel{\label{cont duality 1}
  \d \phi = \star \d B_{(d - 1)}^\vee.
}
Note that the Hodge star maps between original and dual lattices in the underlying microscopic theory. These expressions make sense as path integral variable maps on a $(d + 1)$-dimensional spacetime.

There are two different ways to twist the duality \eqref{2d KW Zk original}. One is
\algns{\label{2d KW Zk twisted 1}
  \eta_\ell \Phi_{\ell_1}\+ \Phi_{\ell_2} &= \Pi_\ell^\vee, \\
  \Pi_v &= W^\vee_v,
}
and promoting the twists to topological gauge fields gives
\algns{\label{2d KW Zk alt 1}
  \Phi_{\ell_1}\+ \Phi_\ell \Phi_{\ell_2} &= \Pi_\ell^\vee,\\
  \Pi_v &= W_v^\vee.
}
The l.h.s.~is a $\Z_K$ scalar theory coupled to topological gauge fields, and the r.h.s.~is a pure $\Z_K$ gauge theory without any singlet constraints. In the continuum, $K \rar \infty$ notation, this is
\bel{
  A^\mu + \del^\mu \phi = \epsilon^{\mu\nu\lambda} F_{\nu\lambda}^\vee.
}
Generalizing as before to arbitrary $d$, this becomes a duality between a gauged zero-form $\phi$ and a $(d - 1)$ form $B_{(d - 1)}^\vee$, given by
\bel{\label{cont duality 2}
  A + \d \phi = \star \d B_{(d - 1)}^\vee.
}
Note that, unlike \eqref{cont duality 1}, the $(d - 1)$-form theory is not in the one-form singlet sector.

The other twist of \eqref{2d KW Zk original} is
\algns{\label{2d KW Zk twisted 2}
  \Phi_{\ell_1}\+ \Phi_{\ell_2} &= \Pi_\ell^\vee, \\
  \Pi_v &= \eta_v^\vee W^\vee_v,
}
which forces the singlet constraint for the matter theory to be $Q = \prod_v \eta^\vee_v$. The twists are promoted to one-form gauge fields just as in \eqref{2d KW alternative 2}. The continuum duality that follows involves a one-form gauge field --- i.e.~a two-form $B^\vee_{\mu\nu}$ --- and in some choice of sign conventions for continuum fields it reads
\algns{
  \del^\mu \phi &= \epsilon^{\mu\nu\lambda} (B_{\nu\lambda}^\vee + F_{\nu\lambda}^\vee),\\
  \phi &= \epsilon^{\mu\nu\lambda} \del_\mu B_{\nu\lambda}^\vee.
}
Compare this to the $d = 1$ case, \eqref{1d KW Zk alt cont}. It is now straightforward to extrapolate and conclude that the KW (or particle-vortex) duality in any $d$ maps a scalar theory of a field $\phi$ to a theory of a $(d - 1)$-form gauge field $B_{(d - 1)}^\vee$, whose one-form symmetry is gauged by coupling to a $d$-form $B_{(d)}^\vee$,
\algns{\label{cont duality 3}
  \d \phi &= \star\big(\d B_{(d - 1)}^\vee + B_{(d)}^\vee\big),\\
  \phi &= \star \d B_{(d)}^\vee.
}

To summarize, the lattice $d = 2$ dualities discussed here were
\boxedAlgns{
  \Z_K\ \trm{clock\ model}/\Z_K &= \Z_K\ \trm{gauge\ theory}^\vee/(\Z_K\ \textrm{one-form}),\\
  \Z_K\ \trm{clock\ model} &= \trm{gauged} \ \Z_K\ \trm{gauge\ theory}^\vee,\\
  \trm{topologically\ gauged}\ \Z_K \ \trm{clock\ model} &= \Z_K \trm{\ gauge\ theory}.
}
The continuum dualities emerging from them are examples of particle-vortex duality:
\boxedAlgns{
  \trm{compact\ scalar/U(1)} &= \textrm{U(1)\ gauge\ theory/(U(1)\ one-form)},\\
  \trm{compact\ scalar} &= \textrm{gauged\ U(1)\ gauge\ theory},\\
  \trm{topologically\ gauged\ compact\ scalar} &= \textrm{U(1)\ gauge\ theory}.
}
As in $d = 1$, there are more dualities not presented here. In particular, orbifolds and their duals can be easily constructed.
\subsection{Fradkin-Kadanoff and its twisting} \label{subsec FK Zk}

The standard parafermion-boson duality in $d = 1$ was presented in \cite{Fradkin:1980th} by Fradkin and Kadanoff (FK) as a direct generalization of the JW map \eqref{1d JW original}. Using conventions from \eqref{def paras} and \eqref{para comm rels}, the FK duality is
\algns{\label{1d FK original}
  \zeta_v &= \Pi_1 \ldots \Pi_{v - 1} \Phi_v,\\
  \xi_v &= \omega^{(K + 1)/2} \Pi_1 \ldots \Pi_{v - 1} \Pi_{v} \Phi_{v}.
}
For $K = 2$, letting $\Pi \mapsto Z$ and $\Phi \mapsto X$, eq.~\eqref{1d FK original} reproduces the JW duality. (Note that, just like in the previous subsection, the roles of position and momentum operators are reversed compared to the original $\Z_2$ duality.) The factor of $\omega^{(K + 1)/2}$ is a convention; to get $\xi_v^K = \1$, it is enough to take any power $\omega^x$ such that $x + \frac12(K - 1) \in \Z$. The choice $x = \frac12(K + 1)$ allows \eqref{1d FK original} to reduce to \eqref{1d JW original} when $K = 2$.

Experience from subsection \ref{subsec 2d JW} suggests that a more local, singlet/singlet version of the FK duality will be more amenable to twisting and generalizing to higher $d$. To this end, consider introducing the parafermion bilinears
\algns{\label{def Pi Sigma}
  \Pi_v &\equiv \omega^{(K + 1)/2} \zeta_v\+ \xi_v, \quad\trm{hence}\quad \Pi_v\+ = \omega^{(K - 1)/2} \xi_v\+ \zeta_v,\\
  \Sigma_{uv} &\equiv \omega^{(K - 1)/2} \zeta_u\+ \xi_v, \quad\trm{hence}\quad \Sigma_{uv}\+ = \omega^{(K + 1)/2} \xi_v\+ \zeta_u.
}
(Recall that $\zeta$ and $\xi$ generalize $\chi$ and $\chi'$, respectively: thus it is $\Pi\+$ and $\Sigma\+$ that are direct generalizations of $Z$ and $S$ from \eqref{def S Z}.) The parafermion hopping operators between adjacent sites, and their FK duals, are
\algns{
  \Sigma_{v + 1, v} &= \Phi\+_{v + 1} \Phi_v,\\
  \Sigma_{v, v + 1} &= \Phi\+_v \Pi_v \Pi_{v + 1} \Phi_{v + 1}.
}
In particular, $\Pi_v$ and $\Sigma_{v + 1, v}$ generate the algebra that commutes with $Q = \prod_v \Pi_v$. They can be used to construct singlet/singlet dualities. The nontrivial commutation relations between these operators are
\gathl{
  \Pi_v \Sigma_{v + 1, v} = \omega^{-1} \Sigma_{v + 1, v} \Pi_v,
  \qquad \Pi_{v + 1} \Sigma_{v + 1, v} = \omega \Sigma_{v + 1, v} \Pi_{v + 1}, \\
  \Pi_v \Sigma_{v, v + 1} = \omega \Sigma_{v,v + 1} \Pi_v,
  \qquad \Pi_{v + 1} \Sigma_{v, v + 1} = \omega^{-1} \Sigma_{v,v + 1} \Pi_{v + 1}.
}

The most apparent singlet/singlet FK duality is thus simply
\algns{\label{1d FK restricted}
  \Sigma_{v + 1, v} \equiv \omega^{(K - 1)/2} \zeta\+_{v + 1} \xi_v &= \Phi\+_{v + 1} \Phi_v,\\
  \omega^{(K + 1)/2} \zeta\+_v \xi_v &= \Pi_v.
}
This generalizes \eqref{1d JW restricted} and assumes that $\zeta_{N + 1} \equiv \zeta_1$. The second line gives the parafermionic expression for the $\Z_K$ symmetry generator (parafermion number mod $K$),
\bel{
  Q = \omega^{N(K + 1)/2} \prod_v \zeta\+_v \xi_v.
}
The first line in \eqref{1d FK restricted} leads, via repeated application of eqs.~\eqref{para comm rels}, to the singlet constraint
\bel{\label{1d para vertex rel 1}
  \1 = \omega^{N(K - 1)/2} \zeta_2\+ \xi_1 \zeta_3\+ \xi_2 \cdots \zeta_1\+ \xi_N = \omega^{N(K + 1)/2 - 1} \zeta_1\+ \xi_1 \cdots \zeta_N\+ \xi_N = \omega^{-1}\, Q,
}
or $Q = \omega \1$ for short.

Two twists of \eqref{1d FK restricted} are possible, as before, but in $d = 1$ they both land on the original FK map \eqref{1d FK original} once the twists go dynamical. In subsection \ref{subsec twisting} the fermionic side was twisted, so just to liven things up, now consider twisting the bosonic side instead:
\algns{
  \Sigma_{v + 1, v} &= \eta_{v + \frac12}^* \Phi_{v + 1}\+ \Phi_v,\\
  \omega^{(K + 1)/2} \zeta\+_v \xi_v &= \Pi_v.
}
The singlet constraint becomes $Q = \omega \prod_v \eta_{v + \frac12}^* \, \1$, or
\bel{
  QW = \omega\1
}
after the twists are promoted into $\Z_K$ gauge fields. The same kind of unusual gauge constraint \eqref{1d flux attached gauge constraint} can now be imposed, and the story is entirely analogous. The $d = 1$ FK dualities are thus
\boxedAlgns{
  \Z_K \ \trm{parafermions} &= \Z_K \ \trm{clock\ model},\\
  \Z_K \ \trm{parafermions}/\Z_K &= \Z_K \ \trm{clock\ model}/\Z_K,\\
  \Z_K \ \trm{parafermions} &= \textrm{flux-attached\ gauged\ } \Z_K \trm{\ clock\ model}.
}
As in the previous subsection, it is further possible to orbifold these models at the expense of introducing $\Z_r$ gauge fields on the other side of the duality. These dualities will not be presented here.
\bigskip

In $d = 2$, duals of parafermion theories on $\Mbb$ are flux-attached $\Z_K$ gauge theories on $\Mbb^\vee$. As parafermions in higher dimensions are not frequently discussed, it may be useful to overview the structure of this theory without reference to dualities. As mentioned above, an absolute ordering of vertices is needed to specify the parafermion operator algebra. This ordering will also induce link orientations: for every link $\ell$, its vertices can be labeled $\ell_{1/2}$ such that $\ell_1 > \ell_2$, and the link will be oriented from $\ell_1$ to $\ell_2$. This orientation choice is not necessary, and it does not agree with the $d = 1$ orientation used above, but it simplifies the following analysis. The parafermion bilinears of interest are $d = 2$ generalizations of \eqref{def Pi Sigma},
\algns{
  \Pi_v &= \omega^{(K + 1)/2} \zeta_v\+ \xi_v,\\
  \Sigma_\ell &= \omega^{(K - 1)/2} \zeta_{\ell_1}\+ \xi_{\ell_2}.
}
As usual, $\Sigma_\ell$ will also be denoted $\Sigma_{\ell_1, \, \ell_2}$. The commutation relations are obtained from eq.~\eqref{para comm rels}, and the most important ones are
\gathl{\label{para bilins comm rels}
  \Pi_{\ell_1} \Sigma_\ell = \omega \Sigma_\ell \Pi_{\ell_1},\qquad
  \Pi_{\ell_2} \Sigma_\ell = \omega^{-1} \Sigma_\ell \Pi_{\ell_2},\\
  \Sigma_{uv} \Sigma_{vw} = \Sigma_{vw} \Sigma_{uv}, \qquad
  \Sigma_{uv} \Sigma_{wv} = \omega^{-\theta(u, w)} \Sigma_{wv} \Sigma_{uv}.
}
where $\theta(u, w) = 1$ if $u > w$, and $\theta(u, w) = -1$ if $u < w$. Just like in the fermionic case, the hopping operators $\Sigma_\ell$ commute if the links do not share a vertex (or, if they do share one, if one flows into it while the other emanates out of it), all operators $\Pi_v$ commute with each other, and the entire algebra generated by $\{\Pi_v, \Sigma_\ell\}$ has a center generated by $Q = \prod_v \Pi_v$, the parafermion number mod $K$. Unlike the ordinary fermion story, the commutation relations of hopping operators involve the curious factor of $\omega^{-\theta(u, w)}$. Its only r\^ole appears to be to alter the flux attachment rules by replacing certain powers of $\omega$ with $\omega^{-1}$. On regular lattices there appears to be no issue due to its presence, but it is not clear if it makes any duality inconsistent on some more generic triangulation.

There is a vertex relation along each face $f \in \Mbb$, namely
\bel{\label{2d para proto vertex rel 1}
  \prod_{i = 1}^{|f|} \Sigma_{v_i v_{i + 1}} = \omega^{-1} \prod_{i = 1}^{|f|} \Pi_{v_i}.
}
Due to the absolute ordering, no one-cycle can have all links oriented the same way relative to the direction of traversal of the cycle. To deal with the resulting $\Sigma$'s going in the ``wrong'' direction, it is useful to note that
\bel{
  \Sigma_{\ell_2 \ell_1} = \omega^{-1} \Sigma_{\ell_1\ell_2}\+ \Pi_{\ell_1} \Pi_{\ell_2} = \Pi_{\ell_1} \Sigma_\ell\+ \Pi_{\ell_2}.
}
Using this identity, eq.~\eqref{2d para proto vertex rel 1} can, up to a power of $\omega$, be written as
\bel{\label{2d para proto vertex rel 2}
  \prod_{\ell \subset f} \Sigma_\ell^{o(\ell; f)} \propto \sideset{}' \prod_{v\subset f} \Pi_v^{o(v; f)}.
}
This is a generalization of the vertex relation \eqref{2d proto vertex relation}, with primes again denoting that the product runs over those sites that join links traversing $\del f$ along the same direction. On the l.h.s.~the signs $o(\ell; f)$ are $+1$ if $\ell$ is oriented counter-clockwise along $\del f$, and $-1$ if $\ell$ is oriented clockwise. On the r.h.s, $o(v; f)$ is $+1$ if $v$ joins links that are both counter-clockwise along $\del f$, and $-1$ if $v$ joins links that are both clockwise.

It should by now be clear that the appropriate generalization of the FK duality \eqref{1d FK restricted} to $d = 2$, and of the JW duality \eqref{2d JW twisted} to $\Z_K$, is
\algns{\label{2d FK}
  \eta_\ell \Sigma_\ell &= \~\Pi_\ell^\vee,\\
  \Pi_v &= W_v^\vee.
}
As before, some background fields $\eta_\ell$ need to be coupled to the parafermions to make sure the vertex relations \eqref{2d para proto vertex rel 2} all map to a consistent Gauss law $\~G^\vee_f = \1$: the background flux $(\delta\eta)_f$ must be chosen to match the proportionality constant in this relation at each $f$. Alternatively, all $\eta_\ell$ can be set to unity but the dual gauge theory must be coupled to a nontrivial background charge density. This form of the duality is consistent with the rule that link orientations on $\Mbb^\vee$ are obtained by rotating those on $\Mbb$ by $\pi/2$ \emph{counter-clockwise}. Note that this happens to be the opposite convention from the one in subsection \ref{subsec KW Zk}.

The rules of flux attachment on $\Mbb^\vee$ (anchor assignment and framing rules $c(\ell)$ of line operators $\~\Pi_\ell^\vee$, cf.~\eqref{def tilde X}) are, as before, mostly fixed by commutation relations, i.e.~by orientation choices on $\Mbb$. For instance, on the square lattice as in footnote \ref{foot sq lattice}, there are still two consistent choices of flux attachment per face.  The $\Z_K$ gauge theory \eqref{2d FK} is in the singlet sector of the anomalous one-form $\Z_K$ symmetry, while the parafermions are in the singlet sector of the zero-form symmetry generated by $Q$. Gauging the latter symmetry lifts the one-form singlet constraint in the gauge theory, as before.

The $K \rar \infty$ limit does not appear particularly interesting on either side of the duality. Parafermions with $K \rar \infty$ are indistinguishable from bosons at states with low parafermion numbers, as $\omega \rar 1$. In other words, they are like compact scalars $\phi$ for all states with much less than $K$ bosons. Similarly, the amount of flux attachment goes to zero in the dual theory, meaning that it will look like a regular U(1) gauge theory in states with low excitations. It would be interesting to understand whether the regime of highly excited states in these theories becomes more tractable at $K \rar \infty$ than at finite $N$.

A more interesting question concerns $\Z_K$ theories with $q$ units of flux attachment when $q \neq \pm 1\, \trm{mod}\, K$. These have Gauss operators
\bel{
  \~G_f^\vee \equiv G_f^\vee \prod_{v:\, v_0 = f} \big(W_v^\vee\big)^q.
}
The gauge-invariant operators here are $W_v^\vee$ and $\~\Pi^\vee_\ell = \Pi^\vee_\ell \big(\Phi_{c(\ell)}^\vee\big)^q$. What are their parafermion duals?

There are two cases to inspect. If $q$ does not divide $K$, the entire gauge-invariant algebra can be generated by $\~\Pi^\vee_\ell$ and $\big(W^\vee_v\big)^q$. Their commutation relations are just like the ones in eq.~\eqref{para bilins comm rels}, except with the substitution $\omega \mapsto \omega^q$. Thus the dual of this theory is a set of parafermions with ``fractional statistics,'' i.e.~with factors of $\omega^q = \e^{2\pi\i q/K}$ instead of $\omega$ appearing in commutation relations \eqref{para comm rels}, and with
\algns{\label{2d FK any}
  \eta_\ell \Sigma_\ell &= \~\Pi_\ell^\vee,\\
  \Pi_v &= \big(W^\vee_v\big)^q.
}
It is therefore possible to find a bosonic dual to \emph{any} theory with statistics governed by a rational number $q/K$ (the \emph{$q/K$ parafermions}). Thus the set of all flux-attached theories is mapped by duality to the set of all  possible Abelian anyons.

If $K = q r$ for $r \in \Z$, $\big(W_v^\vee\big)^q$ and $\~\Pi^\vee_\ell$ satisfy the algebra of $\Z_r$ parafermion bilinears, and indeed the dual theory is that of $\Z_r$ parafermions coupled to $\Z_q$ scalar twist fields. To see this, first note that the gauge theory target space can be understood as the orbifold $\Z_K/\Z_q$: this is a flux-attached $\Z_r$ gauge theory coupled to ``twist fields'' that change from one of the $q$ copies of $\Z_r$ to the next one when the electric field $\Pi_\ell$ tries to shift the field at $\ell$ from $\qvec{\omega^{n r - 1}}$ to $\qvec{\omega^{nr}}$. Thus the $\Z_K$ theory decomposes into a flux-attached $\Z_r$ theory and an ordinary $\Z_q$ gauge theory, with $\~\Pi_\ell^\vee$ understood as a product of the flux-attached electric operator for $\Z_r$ fields and an ordinary electric operator for $\Z_q$ fields that activates only when the $\Z_r$ field is a chosen state, say $\qvec{\omega^{r - 1}}$, on the given link. Each of these can be dualized independently, giving a theory of $\Z_r$ parafermions coupled to scalar $\Z_q$ twist fields. In summary:
\boxedAlgns{
  \Z_K \trm{\ parafermions}/\Z_K &= \textrm{flux-attached} \ \Z_K \ \trm{gauge\ theory}^\vee/(\Z_K\ \textrm{one-form}),\\
  \frac qK\trm{\ parafermions}/\Z_K &= q\textrm{-flux-attached}\ \Z_K\ \trm{gauge\ theory}^\vee/(\Z_K\ \textrm{one-form}).
}

\subsection{Comments on paraspin structures} \label{subsec paraspin structures}

Recall that the discussion of spin structures in subsection \ref{subsec 2d spin structures} was initiated as the discussion of twist fields. It is thus natural to identify the twists of parafermion dualities with \emph{paraspin structures} on the lattice $\Mbb$. These are topological $\Z_K$ fields whose curvature is fixed locally but whose holonomies can be dynamical. If the holonomies are merely background fields, the theory can be called a \emph{paraspin theory}, in analogy with the $\Z_2$ case. If the holonomies are dynamical, the theory is a \emph{non-paraspin} theory. Unlike the $\Z_2$ case, however, here there is no notion of a $\Z_K$ Stiefel-Whitney class, and one may wonder how natural is it to even define paraspin structures.

A conjectural answer to this question is that paraspin structures may naturally appear only in low dimensions. Recall that the universal cover of SO(2), the Euclidean rotation group in $D = 2$, is $\R$. Thus defining $\Spin(2) $ such that $\SO(2) = \Spin(2)/\Z_2$ appears unnatural; one could easily define $\Spin_K(2)$ as the $K$-fold cover of $\SO(2)$ such that $\SO(2) = \Spin_K(2)/\Z_K$. Similarly, in $D = 3$ the representations used to classify massive particles are governed by the little group $\SO(2)$, which can be uplifted to any $\Spin_K(2)$. This matches well with the lore that parafermions are sensible quantum degrees of freedom in $d = 1$, while they may appear as massive excitations (Abelian anyons) in topological theories in $d = 2$. It would be interesting to develop a discrete theory of paraspin structures and $\Z_K$ Stiefel-Whitney classes, and to see in these discrete setups why such objects might be unnatural in higher dimensions. Steps in this direction have been taken in \cite{Geiges:2010, Randal-Williams:2010, Runkel:2018feb}.

\section{Conclusion}

This paper has collected and derived a large number of exact dualities in $d = 1$ and $d = 2$. Some higher dimensional dualities have also been shown. Most of the dualities here are not new, with the notable exception of dualities with fractional statistics in subsection \ref{subsec FK Zk}. Moreover, these are certainly not \emph{all} the exact dualities in low dimensions. Nevertheless, the analysis in this paper may prove to be a useful step towards systematizing our knowledge of dualities in general. Immediate generalizations to be explored involve writing explicit dualities in higher dimensions, understanding the details of what goes wrong when the Stiefel-Whitney class $w_2$ is not exact, and extending this story to include spin$^c$ structures. A longer-term generalization that is also intriguing is the extension of the discussion presented here to any class of nonabelian lattice theories.

Two rather ambitious problems remain open. One concerns explicit discretizations of Chern-Simons theory. The flux-attached gauge theories of subsections \ref{subsec 2d JW} and \ref{subsec FK Zk} have Gauss laws, line operators (anyon statistics), and framing dependence that bring to mind Chern-Simons theories. It is easy to come up with particular Hamiltonians for such theories which break time-reversal symmetry and are described by topological field theories at long distances. Unfortunately, in the simplest models the ground state degeneracies of these theories do not match the Chern-Simons ones, and there are no protected edge modes. These issues do not appear unsurmountable. By further projecting to a part of the Hilbert space, the ground state degeneracy for a $\Z_2$ flux-attached gauge theory on a lattice of genus $g$ can be brought down from $2^{2g}$ to $2^g$ (though whether this is physically interesting remains to be seen). Moreover, by making sure to give the massive excitations a nontrivial band structure (say, by taking the particular $\Z_2$ gauge theory to be dual to a fermionic Chern insulator of some sort), a protected edge spectrum and quantum Hall behavior may be obtained. Understanding the details of such constructions is a task for the (hopefully near) future.

The second problem of interest is the quest for a spin-statistics relation in discrete setups.\footnote{I thank D.~Gaiotto for a discussion on these issues.} Subsection \ref{subsec 2d spin structures} has emphasized that spin structures can indeed naturally appear without the crutch of relativity. These spin structures are crucial ingredients in defining JW dualities, and the corresponding paraspin structures are equally important for FK dualities. These dualities, in turn, are the most direct ways to \emph{define} what one means by a (para)fermionic Hilbert space, as commented below eq.~\eqref{1d JW original}. This is because any many-body system with nontrivial statistics must generically be given a graded Hilbert space, which depends on a choice of ordering of creation operators. By bosonizing these systems, all these ordering ambiguities are transferred into a choice of twisting. (This still does not uniquely fix the bosonic system, cf.~footnote \ref{foot sq lattice}, but the remaining choices on the bosonic side do not influence ordering ambiguities.) By promoting the twists into dynamical variables, these ambiguities are effectively averaged over, resulting in a theory whose Hilbert space can be constructed without obstructions. The general question that remains is this: is it possible to prove that \emph{any} definition of a (para)fermionic Hilbert space --- even without using duality --- necessitates the introduction of a (para)spin structure?

\section*{Acknowledgments}

I have the pleasure to thank Arun Debray, Lorenzo Di Pietro, Paul Fendley, Davide Gaiotto, Simeon Hellerman, Chao-Ming Jian, Theo Johnson-Freyd, Anton Kapustin, Zohar Komargodski, Raghu Mahajan, Tatsuma Nishioka, Kantaro Ohmori, Yuji Tachikawa, Masahito Yamazaki, and Weicheng Ye for stimulating discussions or comments. Research at Perimeter Institute is supported by the Government of Canada through Industry Canada and by the Province of Ontario through the Ministry of Economic Development \& Innovation.

\end{document}